\documentclass[a4paper,12pt,oneside]{report}
\pdfoutput=1

\usepackage{lmodern}
\usepackage{float}
\usepackage{textcomp}
\usepackage{gensymb}
\usepackage{amsmath}
\usepackage[english]{babel}
\usepackage[utf8]{inputenc}
\usepackage[babel]{csquotes}
\usepackage{sectsty}
\usepackage[toc,page]{appendix}
\usepackage{chngcntr}
\usepackage{url}

\usepackage{lscape}                                                    
\allsectionsfont{\raggedright}
\usepackage[pdftex]{graphicx}
\graphicspath{ {./anc/} }
\usepackage{epstopdf}
\counterwithin{figure}{section}
\usepackage{subcaption}                                                    
\usepackage{tikz}                                                        

\tikzset{font={\fontsize{12pt}{12}\selectfont}}                            
\usepackage{caption}
\usepackage[final]{pdfpages}

\usepackage{booktabs}
\usepackage{multirow}
\usepackage{rotating}

\usepackage[nottoc,notlot,notlof]{tocbibind}
\usepackage[hypertexnames=false]{hyperref}
\usepackage{bookmark}
\usepackage[all]{hypcap}                
\hypersetup
{
    pdfauthor = {author},
    pdfsubject = {subject},
    pdftitle = {title},
    pdfkeywords = {kw},
    hidelinks,
    linkbordercolor = {white},
}
\usepackage{tablefootnote}
\begin{document}

\begin{titlepage}
\begin{center}

\newcommand{\HRule}{\rule{\linewidth}{0.5mm}}

\begin{center}
    \includegraphics[width=0.9\linewidth]{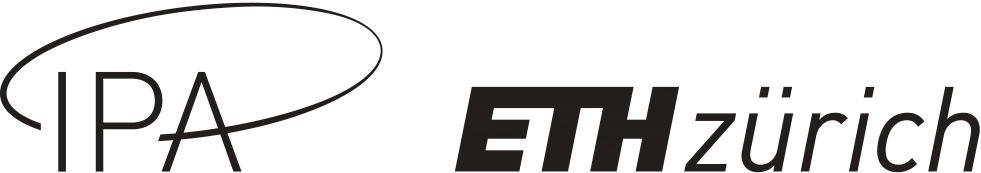}
\end{center}

\vspace*{2cm}

\HRule \\[0.4cm]
{ \huge \bfseries Monitoring the optical quality of the FACT Cherenkov Telescope\\[0.4cm] }
\HRule \\[1.5cm]

{\large \bfseries Master's Thesis}\\[1cm]

{\bfseries \Large Laurits Tani}\\[2cm]

\begin{minipage}[t]{0.4\textwidth}
    \hspace*{\fill}Supervisor:
\end{minipage}
\hspace*{2mm}
\begin{minipage}[t]{0.4\textwidth}
    Prof. Dr. Adrian Biland
\end{minipage}

\vfill

{\large 2019}

\end{center}
\end{titlepage}


\pdfbookmark[section]{Abstract}{abs}
\chapter*{Abstract}
    In gamma ray astronomy muon events have a distinct feature of casting ring-like images on the sensor plane, thus forming a well known signal class for Cherenkov telescopes. These ring-like images can then be used to deduce the optical point spread function (PSF) which is an important measure of the optical quality of the imaging-reflector. In this thesis the observed 'fuzziness' of muon rings is used as a measure to infer the PSF. However to have a good estimate for this 'fuzziness' parameter, the reconstruction of the ring center and ring radius itself needs to be accurate, so different methods of ring feature extraction are studied. To check for the accuracy of the methods a simulation and analysis is performed. Measuring the evolution of the PSF over time allows to identify its effects and take them into account for the reconstruction of gamma-rays postliminary. As a further benefit of the methods presented here no additional observations are needed to measure the PSF nor any human activity on site is required. The accuracy of the method, and the PSF of FACT vs. time are presented.
\thispagestyle{empty}

\newpage

\tableofcontents
\thispagestyle{empty}

\newpage
\addtocontents{toc}{\protect\thispagestyle{empty}}

\chapter{\label{sec:Intro}Introduciton}
    \section{Astroparticle Physics}
        Astroparticle physics investigates the origins of particles of astronomical origin. Insight into the structure and history of our universe is gained by researching the mechanisms responsible for e.g. accelerating particles to high energies, origins of charged cosmic rays and the nature of dark matter.

        \subsection{Astrophysical messengers}
            Particles coming from all around the Universe that are constantly bombarding Earth are used as messengers to explore the cosmos. These particles can be divided into three classes: (electrically charged) cosmic rays, neutrinos and photons.

            \begin{figure}[H]
                \begin{center}
                    \includegraphics[width=0.7\linewidth]{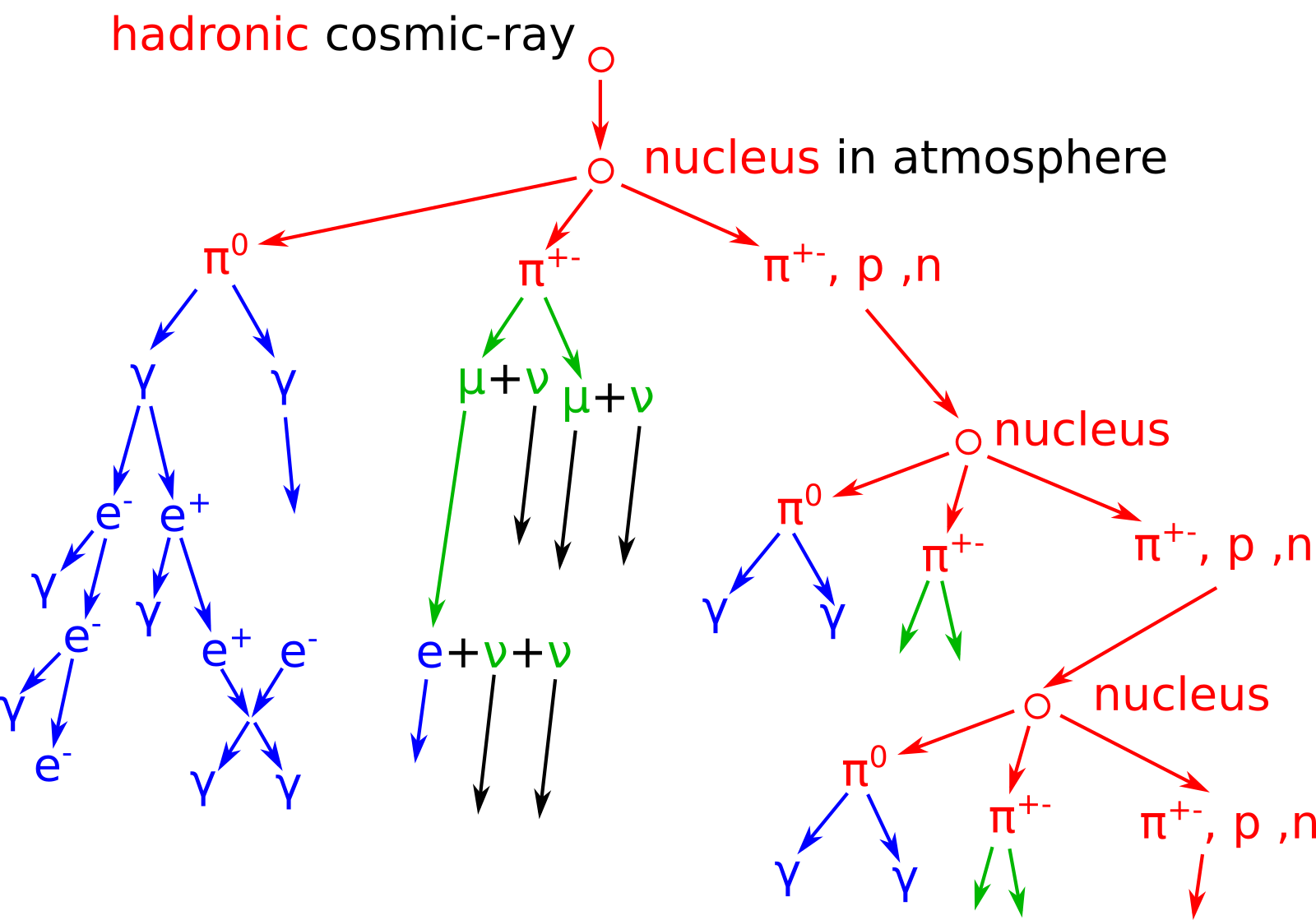}
                    \caption{Hadronic shower \cite{mueller2019cherenkov}}
                    \label{fig:airshower}
                \end{center}
            \end{figure}

            \subsubsection{(Charged) Cosmic Rays}
                Cosmic rays consist mainly ($\sim$ 90$\%$) out of protons. Upon impact with the Earth's atmosphere, they can produce showers of secondary particles that in case of high enough energy will also reach the ground. This class of airshowers is called hadronic airshower (see figure \ref{fig:airshower}). Despite being the easiest to detect with Cherenkov teleskopes, they do not carry information about their origins. The energy spectrum of cosmic rays is very featureless, though there are 3 features: two knees and ankle (respectively at energies $\sim 10^{15}$ eV, $\sim 10^{17}$ eV and $\sim 10^{19}$ eV). \cite{olive2014review}

                However cosmic rays do have a theoretical upper limit for energy. It is called Greizen-Zatsepin-Kuzmin limit (or GZK-cutoff for short). The limit comes from the fact that universe is not transparent anymore for most energetic cosmic rays ($\sim 10^{21}$ eV) because they interact with the very low-energetic cosmic microwave background (CMB) photons.

                \begin{equation*}
                    p + \gamma_{CMB} \rightarrow \Delta^{+} \rightarrow n + \pi^{+}
                \end{equation*}
                or
                \begin{equation*}
                    p + \gamma_{CMB} \rightarrow \Delta^{+} \rightarrow p + \pi^{0}
                \end{equation*}

        \begin{figure}[H]
            \begin{center}
                \includegraphics[width=0.9\linewidth]{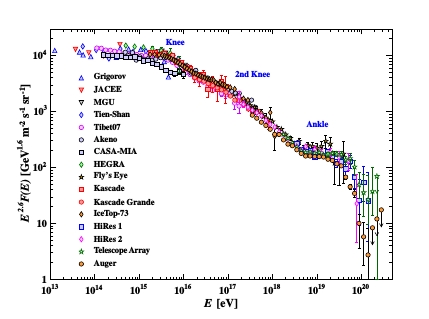}
                \caption{Charged cosmic-ray energyspectrum \cite{olive2014review}}
                \label{fig:CR_energy}
            \end{center}
        \end{figure}

            \subsubsection{Neutrinos}
                Neutrinos are uncharged leptons with a very small cross section, thus being highly uninteractive. In addition to gravity which we can neglect due to almost zero mass of the neutrino, neutrinos interact with matter only via weak interactions - charged current and neutral current. Neutrinos are often times called the ultimate messengers of the universe, because their trajectories are not bent by cosmic magnetic fields, their signal will not get attenuated due to their very uninteractive character, thereby being highly penetrating thus allowing us to see into the core of their origin. Unfortunately being that uninteractive also poses a problem if one were to detect them. For neutrinos to interact with detector material, large detectors are needed. Otherwise they would only fly through the detector without interacting. In interactions they produce leptons ($e^{\pm}, \mu^{\pm}, \tau^{\pm})$, which then produce Cherenkov photons that are easy to detect. As examples for neutrino observatories, there are IceCube and Super Kamiokande. In those experiments the main background is the atmospheric neutrinos.

            \subsubsection{Photons}
                Photons are uncharged elementary particles that have no rest mass. They play a major role in astroparticle physics due to the fact that their trajectories are not bent by the intergalactic/cosmic magnetic fields, thus they carry the information of the place of their origin. Photons that are of interest to us are highly energetic and are called $\gamma$-rays. The detected photon energies cover energy range of more than twenty orders of magnitude from $\sim1$ $\mu$eV (radio waves) up to several hundred TeV ($\gamma$-rays).

        \section{Gamma-ray astronomy}
            Gamma-ray astronomy is the observation of the previously mentioned gamma rays, the most energetic electromagnetic radiation that has energies above 100 keV. $\gamma$-rays originating from the solar flares have energies mostly in the MeV range, although they can be created also in the GeV range. \cite{gammarays} There are several mechanisms to produce these highest energetic photons - inverse Compton effect, annihilation of electron and positron, decay of radioactive material (also known as gamma decay) or the decay of high energetic neutral pion. The highest photon energies recorded so far are in the TeV range.

            The detection of $\gamma$-rays is diffucult, since they are much more rare than the low-energetic X-rays. For gamma-ray energies higher than 100 GeV the flux is so rare, that the effective area of space-bound experiments is too small to have any statistical significance. To overcome this, ground-based telescopes are used, that use the atmosphere as calorimeter. In contrast to charged cosmic-rays, gamma-rays produce only electromagnetic airshowers, that are then detected by ground-based telescopes. The cascade arises from pair-production and Bremsstrahlung of the electrons and positrons. Electrons and positrons that move faster than light in the atmosphere, will also emit Cherenkov radiation.

            Although the Imaging Atmospheric Cherenkov Teleskope techniques have currently the highest sensitivity, $\gamma$-ray astronomy is still limited by the fact, that at low energies non-gamma-ray background dominates and at higher energies the $\gamma$-ray flux is too low.

    \section{FACT Cherenkov Telescope}
        The first G-APD Cherenkov Telescope (FACT) is a ground based Imaging Atmospheric Cherenkov Telescope (IACT) that started operating in October 2011. It is the first IACT to test novel Geiger-mode avalance photo diodes (G-APDs) as photosensors. The focal length of the telescope is 4.889 m and field of view is 4.5\degree, thus for each individual pixel the field of view is $\sim 0.11\degree$ \cite{anderhub2013design}

        The telescope is located next to two 17m MAGIC telescopes at the Observatorio del Roque de los muchachos on the Canary Island La Palma in Spain at the altitude $\sim 2200$ m. It constantly monitors several bright sources, among them for example Mrk421, Mrk501 and the Crab nebula.

        Imaging atmospheric-Cherenkov method is an indirect measurement technique, where Cherenkov photons emitted by charged particles are used for imaging. There are 1440 photosensors, each equipped with 3600 cells, with individual readouts in the camera to detect these photons. Using silicon photo multipliers (SiPM's which is a different name for G-APD) allows observations also under bright light conditions.

        The reflector is comprised of 30 hexogonal mirrors, with the total reflective surface of 9.51 m$^2$. These 30 mirrors are arranged in the Davies-Cotton arrangement and placed at their focal distance to their focal point and oriented in the direction of a point at twice the focal length. \cite{anderhub2013design} \cite{biland2014calibration}

        \begin{figure}[H]
            \begin{center}
                \includegraphics[width=0.7\linewidth]{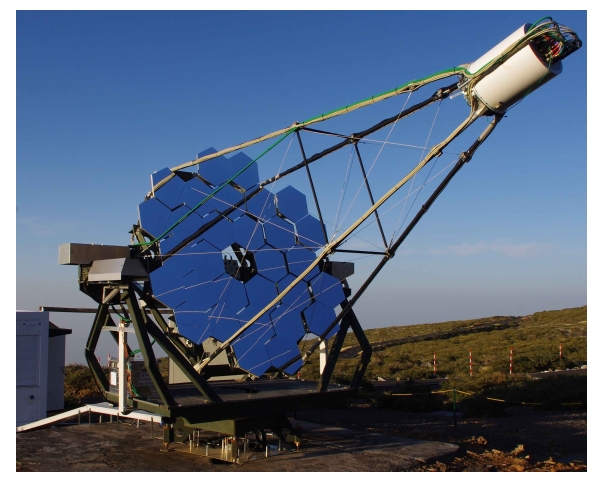}
                \caption{Photo of the FACT telescope at the Observatorio del Roque de los muchachos on La Palma (Canary Islands, Spain) \cite{anderhub2013design}}
                \label{fig:FACT}
            \end{center}
        \end{figure}

        \subsection{Photon-stream representation}
            We describe incoming photons as a list of arrival times for each pixel. This list is called "photon-stream". Due to low electronics-noise and fast read-out of the SiPM sensors, FACT is able to provide the accuracy to take full advantage of the "photon-stream". \cite{photon_stream} \cite{mueller2019cherenkov}

\begin{figure}[H]
            \begin{center}
                \includegraphics[width=1\linewidth]{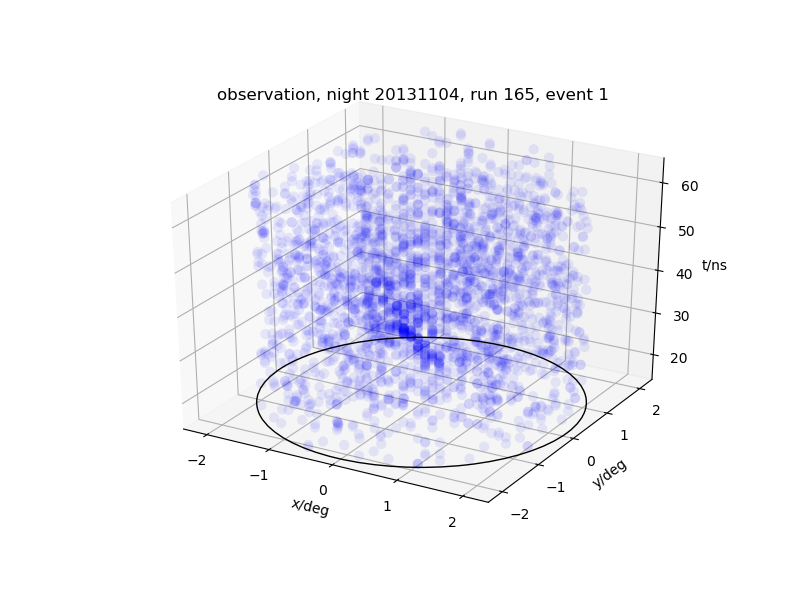}
                \caption{Recorded event in photon-stream representation. Each blue dot corresponds to a single photon.}
                \label{fig:photon_stream}
            \end{center}
        \end{figure}

            On figure \ref{fig:photon_stream} one can see the depiction of an event in that representation.

    \section{Creation of muon rings}
        \subsection{Cherenkov effect}
            Charged particles travelling in a medium with velocities larger than the local speed of light emit Cherenkov photons. These photons are emitted in spherical waves, thus forming a cone of photons with an opening angle that is dependent on the refractive index of the medium and the velocity of the charged particle.

\begin{figure}[H]
                \begin{center}
                    \includegraphics[width=1\linewidth]{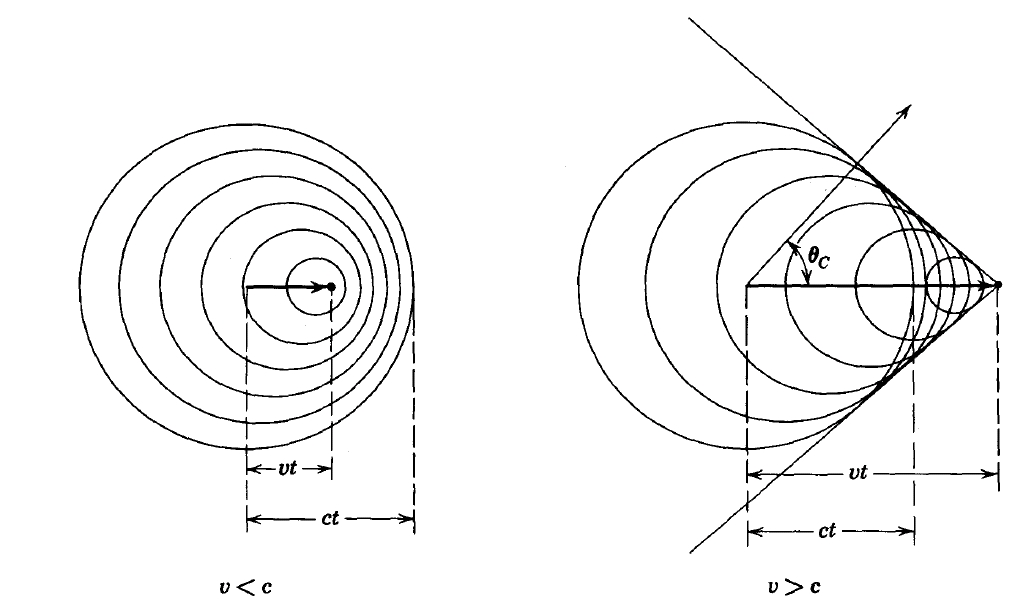}
                    \caption{Cherenkov effect. On the left: charged particle travelling in medium with speed less than lightspeed in that medium ($v<c$). On the right: charged particle travelling in medium with speed larger than lightspeed in that medium ($v>c$). Figure taken from \cite{cherenkov_radiation_picture}}
                    \label{fig:cherenkov_radiation}
                \end{center}
            \end{figure}

            As can be seen on right side figure of figure \ref{fig:cherenkov_radiation}, the opening angle of the cone is:
            \begin{equation} \label{eq:cone_oa}
                \theta_C = arccos\left(\frac{1}{n\cdot\beta}\right)
            \end{equation}
            The refractive index of air at altitude 2200 m is $n = \frac{c_0}{c} = 1 + \epsilon = 1.00022$ where $c_0$ is the speed of light in vacuum, c is the speed of light in medium, and $\epsilon$ is the fractional refractive index.

            In order to produce Cherenkov emission, the following relation must hold:
            \begin{gather}
                \begin{align}
                    v &> c \\
                    \implies \beta & > \frac{1}{n} \\
                    \implies \beta_{min} & = 0.99977
                \end{align}
            \end{gather}

        \subsection{Muons and their properties}
            Muons have a mass of $ m = 105.6583745 \pm 0.0000024$ MeV and a mean lifetime in their reference frame of $ \tau = 2.1969811 \pm 0.0000022 \cdot 10^{-6}$ seconds. Muons are produced in the atmosphere in hadronic airshowers as the decay product of charged pions:

            \begin{equation}
                \pi^- \to \mu^- + \overline{\nu}_{\mu} \nonumber
            \end{equation}
            and
            \begin{equation}
                \pi^+ \to \mu^+ + \nu_{\mu} \nonumber
            \end{equation}

             Muons produced in hadronic showers are usually relativistic particles, thus many reach ground before decaying due to time dilation. The decay products of muons are:
             \begin{equation}
             \mu^- \rightarrow e^{-} + \overline{\nu_e} + \nu_\mu
             \end{equation}
             or
             \begin{equation}
             \mu^{+} + \rightarrow e^+ + \overline{\nu_\mu} + \nu_e
             \end{equation}

            Since the Lorentz factor is defined as
            \begin{equation}
                \gamma = \frac{1}{\sqrt{1-\beta^2}} = \frac{E}{E_0} > 1
            \end{equation}
            then we find that the minimal energy for muons to produce Cherenkov radiation is:
            \begin{equation}
                E_{min} = \frac{E_0}{\sqrt{1-\beta^2_{min}}} \approx 4.93 GeV
            \end{equation}

            Since with higher energy the opening angle increases then we can find out the maximal opening angle corresponding to the maximum speed of the particle ($\beta = 1$). Using equation \ref{eq:cone_oa} we find:

            \begin{equation}\label{eq:openingAngle}
                \theta_{max} = \frac{1}{n} \approx 1.22\degree
            \end{equation}

        \subsection{Muon rings}
            Photons that are hitting the aperture of the telescope are reflected to the sensor. Since the aperture is an imaging mirror, then all photons with the same direction-vector are reflected onto the same place in the sensor. Thus all Cherenkov photons that are emitted in a cone with a certain fixed angle will form a ring-like shape in the sensor plane. Unfortunately muon events are very dim, so they are difficult to detect. For FACT they are among the dimmest events that can be triggered.

            \subsubsection{Zero inclination angle and distance within aperture radius}
\begin{figure}[H]
                    \begin{center}
                        \includegraphics[width=1\linewidth]{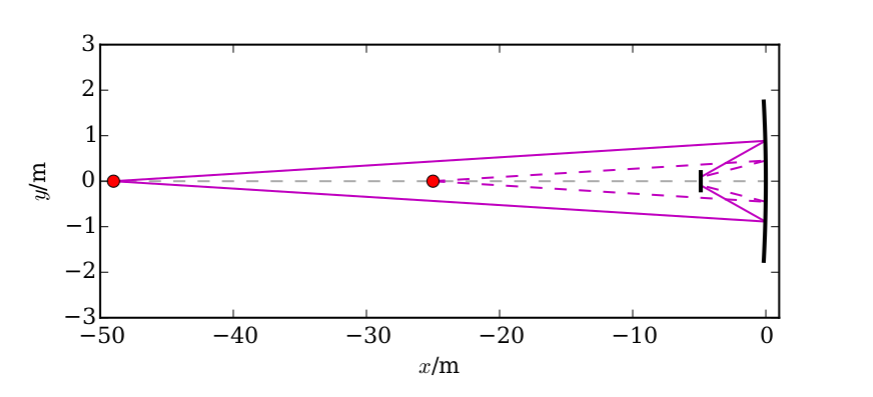}
                        \caption{Formation of the muon ring in case of $\alpha = 0 $ and $r = 0$. Cherenkov cone opening angle $\theta_C = \theta_{max}$ Figure taken from \cite{nothe2014extraktion}}
                        \label{fig:best_ring}
                    \end{center}
                \end{figure}

                In this case muon rings that are forming on the sensor plane will be evenly illuminated and always form in the center of the camera. On figure \ref{fig:best_ring} one can see how muon rings formation in case of zero inclination angle and with an impact radius of zero.

            \subsubsection{Non-zero inclination angle}
\begin{figure}[H]
                    \begin{center}
                        \includegraphics[width=1\linewidth]{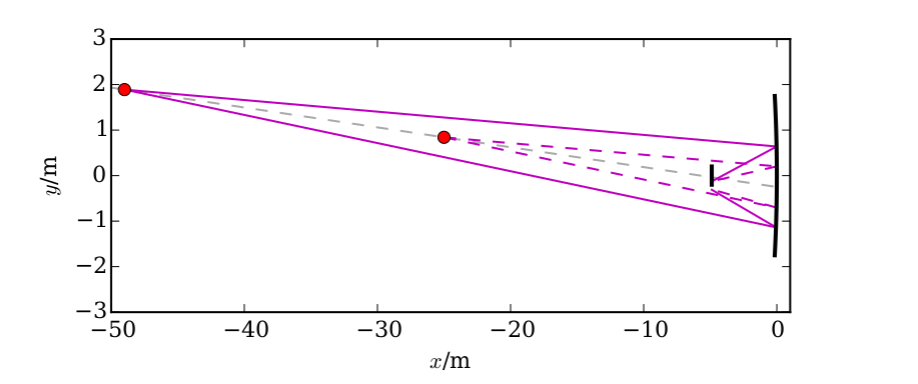}
                        \caption{$\alpha = 2.5^o$, $r = -0.25$m,  $\theta_C = \theta_{max}$. Figure taken from \cite{nothe2014extraktion}}
                        \label{fig:non_zero_inclination}
                    \end{center}
                \end{figure}

                On figure \ref{fig:non_zero_inclination} one can see the case when the inclination angle of the muon is not zero. Depending on the inclination angle of the muon, the center position of the ring moves on the sensorplane.

            \subsubsection{Greater distance than the aperture radius}
\begin{figure}[H]
                    \begin{center}
                        \includegraphics[width=1\linewidth]{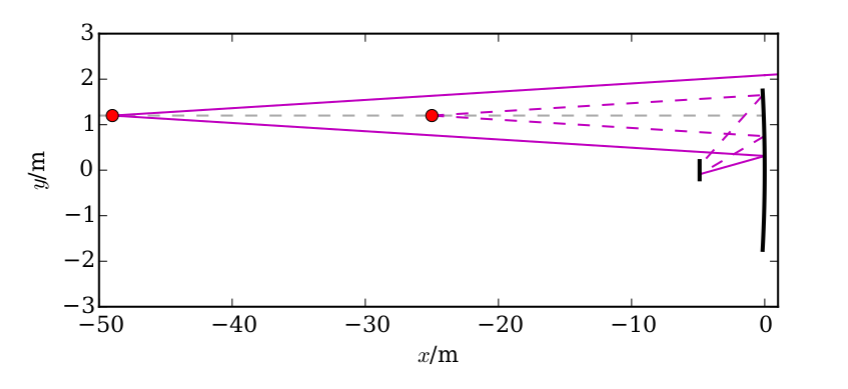}
                        \caption{$\alpha = 0$, $r = 1.2$ m,  $\theta_C = \theta_{max}$ Figure taken from \cite{nothe2014extraktion}}
                        \label{fig:big_impact_parameter}
                    \end{center}
                \end{figure}

                If muons are hitting the aperture plane further away than the radius of the aperture, then the Cherenkov photons do not produce full rings anymore - some amount of photons, depending on the distance, won't hit the mirror anymore. Consequently the muon ring is not illuminated evenly and the ring might be incomplete. On figure \ref{fig:big_impact_parameter} one can see the case when muon is not hitting the aperture of the telescope anymore.

\newpage
\chapter{Simulation}
    In order to benchmark the effectiveness of muon detection and ring feature extraction methods, we simulate FACT muon event observations. For this purpose a custom muon simulation was created. Having a simulation allows us to relate observed 'fuzziness' (the spread of photons around ring line) of the muon ring with the optical point spread function. (For scripts see \cite{MRS})

    \section{Individual muon event}

        A single muon will be simulated, given the following parameters:

        \begin{itemize}
            \item direction vector of the muon
            \item support position of the muon
            \item night-sky-background rate
            \item opening angle of the Cherenkov cone
            \item Cherenkov photon emission rate
            \item point spread function of the reflector
            \item arrival time standard deviation of Cherenkov photons
        \end{itemize}

\begin{figure}[H]
            \begin{center}
                \includegraphics[scale=0.7]{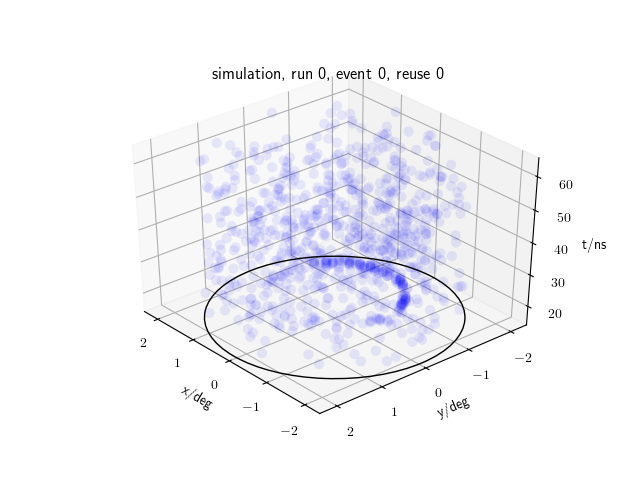}
                \caption{Example of a muon event with $\alpha \neq 0$ and $r_{impact} > r_{aperture}$ and night-sky-background rate of $35\times10^{6} s^{-1}$ which corresponds to the dark night}
                \label{fig:simulation}
            \end{center}
        \end{figure}

        A muon will emit Cherenkov photons when traversing to the ground. Distance travelled before emitting another Cherenkov photon is Poisson distributed. In case the emitted Cherenkov photons hit the aperture plane within the aperture radius, they are reflected onto the sensor plane. Also additional night-sky-background photons are simulated. On figure \ref{fig:simulation} is a simulated muon event with non-zero inclination angle and impact parameter that is larger than aperture radius.

    \section{Multiple muon events}

        \subsection{Choice of parameters} In order to have all possible muon events within a given range muon directions on the skydome and its support position are drawn randomly from uniform distribution. In addition also the opening angle of the Cherenkov cone is drawn from an uniform distribution. These parameters are then given to the function that simulates all these individual muon events.

        \subsection{Simulation process} There are 13 possible input parameters:

            \begin{itemize}
                \item Number of muons to be simulated 
                \item Maximum inclination angle
                \item Maximum aperture radius of muon support vector
                \item Minimum opening angle of the Cherenkov cone [default: 0.4 degrees] 
                \item Maximum opening angle of the Cherenkov cone [default: 1.6 degrees] 
                \item Night-sky-background-photon rate per pixel [default $35\times 10^6 s^{-1} pixel^{-1}$ which corresponds to dark nights] \cite{biland2014calibration}
                \item Standard deviation of the arrival times of the photons [default: $0.5 \times 10^{-9}$ seconds]
                \item Rate of Cherenkov photons to be generated per meter [default: 3.0] (This value was chosen based on observed muon events. Otherwise Frank-Tamm formula, refractive index, mirror reflectivity, photo-sensor-efficiency and possibly even more should be taken into account to get this value.) \cite{frank1991coherent}
                \item FACT aperture radius [default: 1.965 m]
                \item Standard deviation of the Point Spread Function (PSF) [default: 0] (for perfect imaging)

            \end{itemize}

\newpage
\chapter{Methods of investigation} \label{chap:investigation}
    \section{Cherenkov photon detection}
        Due to the fact that muon events are very dim, the biggest obstacle in seeing these is the triggering the telescope. In case of high night-sky-background rate the trigger threshold is higher and the dim muon event will not trigger the telescope.

        Thus the next step is the identification of Cherenkov photons (detection for short) from night-sky-background photons. Because Cherenkov photons reside much more densly than night-sky-background photons, we need to use a density based algorithm like Two-Level-Time-Neighbor-Cleaning or density clustering.

        \subsection{Spatial clustering}
            Spatial clustering is a technique used in data mining, which groups objects into classes or clusters based on their similarities within the cluster. One example of these spatial clustering techniques is DBSCAN (\cite{ester1996density}), that will be used in this thesis.

            \subsection{DBSCAN (DB)}
                DBSCAN (Density-Based Spatial Clustering of Applications with Noise) recognizes Cherenkov photon clusters by taking into account the fact that within this cluster the typical density of photons is much higher than outside of the cluster. Since it can discover arbitrary shaped Cherenkov clusters, it is very powerful. Basic ideas:
                \begin{itemize}
                    \item The neighbourhood (on the $c_x$ and $c_y$ (the directional cosines) plane) that is within a range $\epsilon$ of a given photon is called the $\epsilon$\textit{-neighborhood}
                    \item For a photon to be considered to be the \textit{core point} of a cluster, there has to be a minimum number of photons, \textit{MinPts}, inside its $\epsilon$\textit{-neighborhood}
                    \item A photon is \textit{directly density reachable} if it is within $\epsilon$\textit{-neighborhood} of a core photon
                    \item For a photon to be \textit{density reachable} from photon P to photon Q, there has to be a chain of photons beween that are \textit{directly density reachable}
                    \item For a photon P to be \textit{density-connected} to photon Q both have to to be \textit{density reachable} with respect to a photon O
                    \item All photons that are not inside a density based cluster are considered to be noise.
                \end{itemize}

                A new cluster is created if a photon P has within its $\epsilon$\textit{-neighborhood} more than \textit{MinPts} photons. During iteration some clusters may be connected and lasts until no more photon can be added to any clusters. \cite{borah2004improved}

            \subsection{DBSCAN's effectiveness in detecting Cherenkov photons}
                A benchmark for the sensitivity and precision of this method was set by simulating muon events with a set night-sky-background (corresponding to dark night in this case).

                \bigskip
                Sensitivity (\cite{powers2011evaluation}) is calculated as:

                \begin{equation} 
                     \frac{true\_positives}{true\_positives + false\_negatives}
                \end{equation}

                and precision (\cite{powers2011evaluation}) is calculated as:

                \begin{equation}
                    \frac{true\_positives}{true\_positives + false\_positives}
                \end{equation}

                where:
                \begin{itemize}
                \item \textbf{true\_positives} is the number of photons that were classified correctly to be Cherenkov photons,

                \item \textbf{false\_positives} is the number of photons that were night-sky-background photons that were classified to be Cherenkov photons,

                \item \textbf{true\_negatives} is the number of photons that were classified correctly to be night-sky-ackground photons,

                \item \textbf{false\_negatives} is the number of photons that were Cherenkov photons that were classified to be night-sky-background photons.
                \end{itemize}
                \bigskip
                \bigskip
                The used parameters for DBSCAN were:
                  \begin{center}
                    \begin{tabular}{l l}
                      Parameter & Value\\
                      \hline
                      min-samples & 20\\
                      $\epsilon$-radius & 0.45 degrees\\
                    \end{tabular}
                  \end{center}
                With these parameters the found sensitivity and precision corresponding to dark night night-sky-background were $98.9 \pm 0.3$ and $82.7 \pm 0.4$ respectively.

\begin{figure}[H]
                    \begin{center}
                        \includegraphics[width=0.7\linewidth]{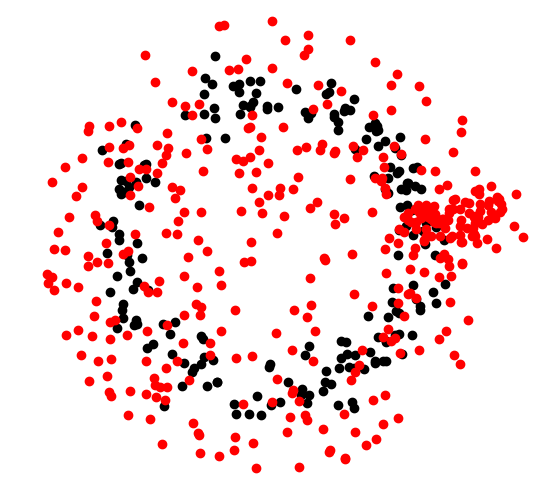}
                        \caption{Inside Cherenkov cluster (black dots) reside also some night-sky-background photons (red dots). Before clustering}
                        \label{fig:cluster_with_nsb}
                    \end{center}
                \end{figure}

                Naturally it is expected that the precision of detecting muons can not be 100 percent due to the fact that night-sky-background photons also will reside in close proximity to the cluster center and thus be considered to part of the cluster (for 2D projection of an event see figure \ref{fig:cluster_with_nsb}). This happens even with the night-sky-background rate corresponding to dark night. 
                Furthermore, detecting Cherenkov photons takes only into account the spatial positioning and time of an individual photon, thus in order to increase precision one needs to apply some other cuts after the density clustering.

        \subsection{DBSCAN's effectiveness with different night-sky-background rate}

\begin{figure}[H]
                \begin{center}
                    \includegraphics[width=0.7\linewidth]{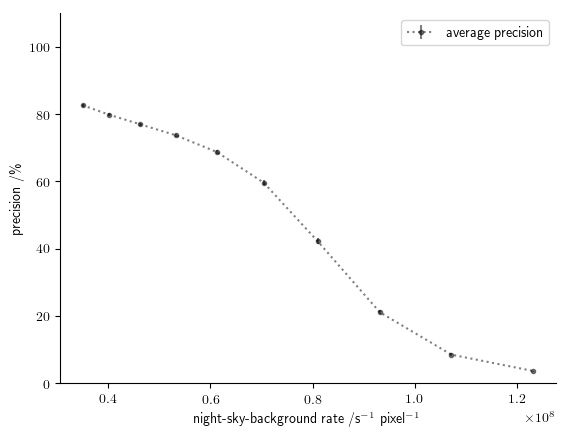}
                    \caption{Precision of the DBSCAN algorithm with different night-sky-background rates}
                    \label{fig:db_prec_noCut}
                \end{center}
            \end{figure}

\begin{figure}[H]
                \begin{center}
                    \includegraphics[width=0.7\linewidth]{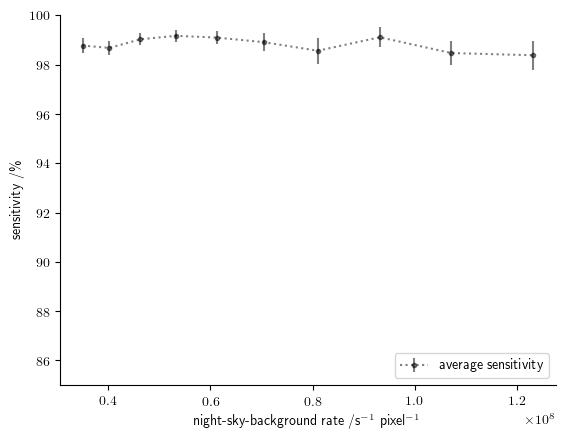}
                    \caption{Sensitivity of the DBSCAN algorithm with different night-sky-background rates}
                    \label{fig:db_sens_noCut}
                \end{center}
            \end{figure}

            Since observations are done under different night-sky-bacground-light-conditions 10 different night-sky-background rates were simulated. Although adjusting DBSCANs parameters might improve the precision, it was not implemented. Regardless one can see from figures \ref{fig:db_prec_noCut} and  \ref{fig:db_sens_noCut} that only precision drops with bigger night-sky-background rate while sensitivity stays rouhly the same. This was of course expected, since increasing the night-sky-background rate only increases the chances of a photon residing in the vicinity of the actual Cherenkov photon cluster and thus being included as one of the elements. At rougly 2 times the dark night rate (dark night night-sky-background rate taken to be $35 \times 10^6$ photons per second per pixel) the precision already drops to roughly 60\%. 

\begin{figure}[H]
                \begin{center}
                    \includegraphics[width=0.6\linewidth]{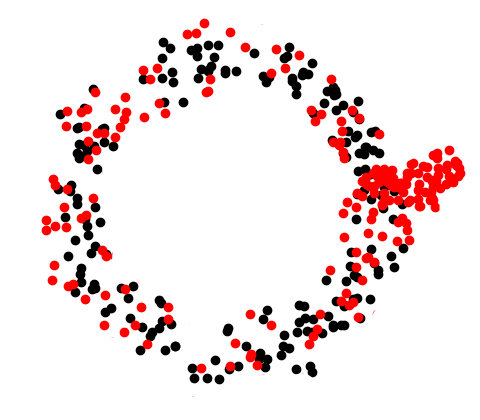}
                    \caption{Having much night-sky-background might cause branches consiting out of night-sky-background photons (red cluster on figure)}
                    \label{fig:branch_example}
                \end{center}
            \end{figure}

            Due to inhomogeneities in the night-sky-background rate and cosmic rays, branches as depicted on figure \ref{fig:branch_example} can emerge due to the fact that only densities are taken into account when searching for Cherenkov photons.

\begin{figure}[H]
                \begin{center}
                    \includegraphics[width=0.7\linewidth]{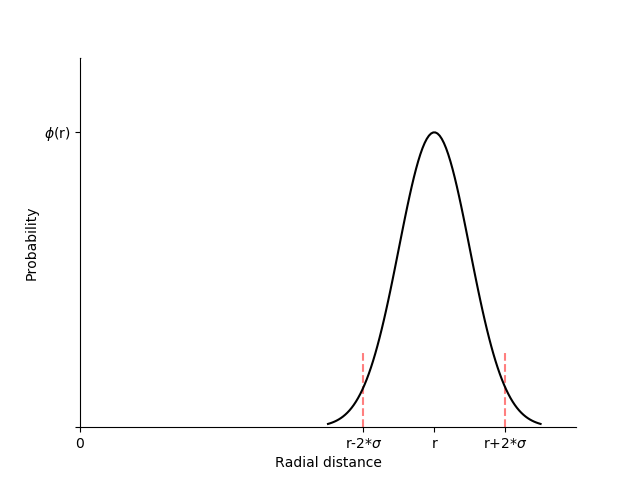}
                    \caption{Implementing cut on the $c_x$-$c_y$ plane (DBSCAN+) when the distance of a photon from reconstructed ring radius is larger than 2 standard deviation}
                    \label{fig:cut_mechanics}
                \end{center}
            \end{figure}

\begin{figure}[H]
                \begin{center}
                    \includegraphics[width=0.5\linewidth]{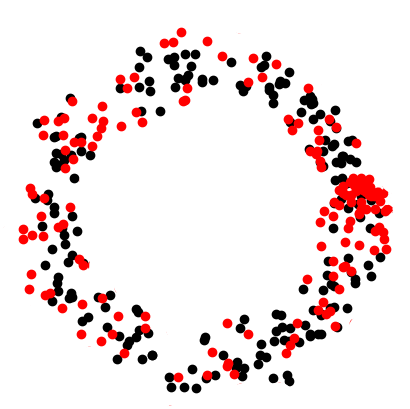}
                    \caption{Implementing cut on the $c_x$-$c_y$ plane (DBSCAN+) when the distance of a photon from reconstructed ring radius is larger than 2 standard deviation}
                    \label{fig:DBSCAN_cut}
                \end{center}
            \end{figure}

            Taking into account muon ring geometry, we can implement a cut, discarding photons that reside from the reconstructed ringline more than 2 standard deviations away (See figure \ref{fig:cut_mechanics}), thus ending up with a muon ring similar to the one on the figure \ref{fig:DBSCAN_cut}.

\begin{figure}[H]
                \begin{center}
                    \includegraphics[width=0.7\linewidth]{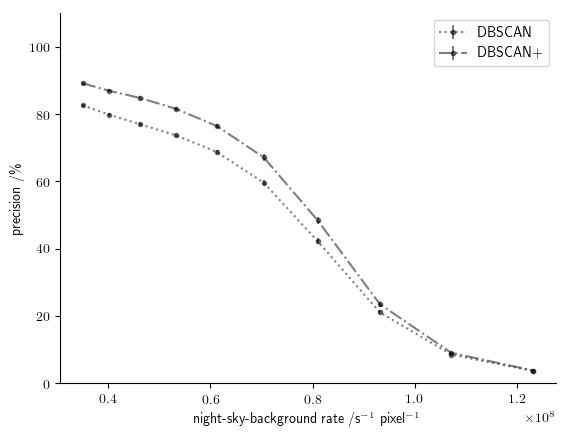}
                    \caption{Precisions of the DBSCAN algorithm with and DBSCAN+ different night-sky-background rates}
                    \label{fig:db_prec_both}
                \end{center}
            \end{figure}

\begin{figure}[H]
                \begin{center}
                    \includegraphics[width=0.7\linewidth]{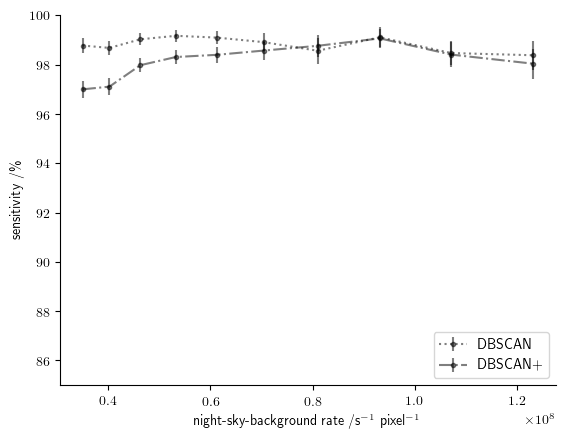}
                    \caption{Sensitivities of the DBSCAN algorithm and DBSCAN+ with different night-sky-background rates}
                    \label{fig:db_sens_both}
                \end{center}
            \end{figure}

            This cut (DBSCAN+) notably increased the precision of the detection while almost not decreasing sensitivity at all as one can see on figures \ref{fig:db_prec_both} and \ref{fig:db_sens_both}. With DBSCAN+ at 2 times the dark night night-sky-background rate the precision has risen to $\sim 70\%$.

\begin{figure}[H]
                \begin{center}
                    \includegraphics[width=\linewidth]{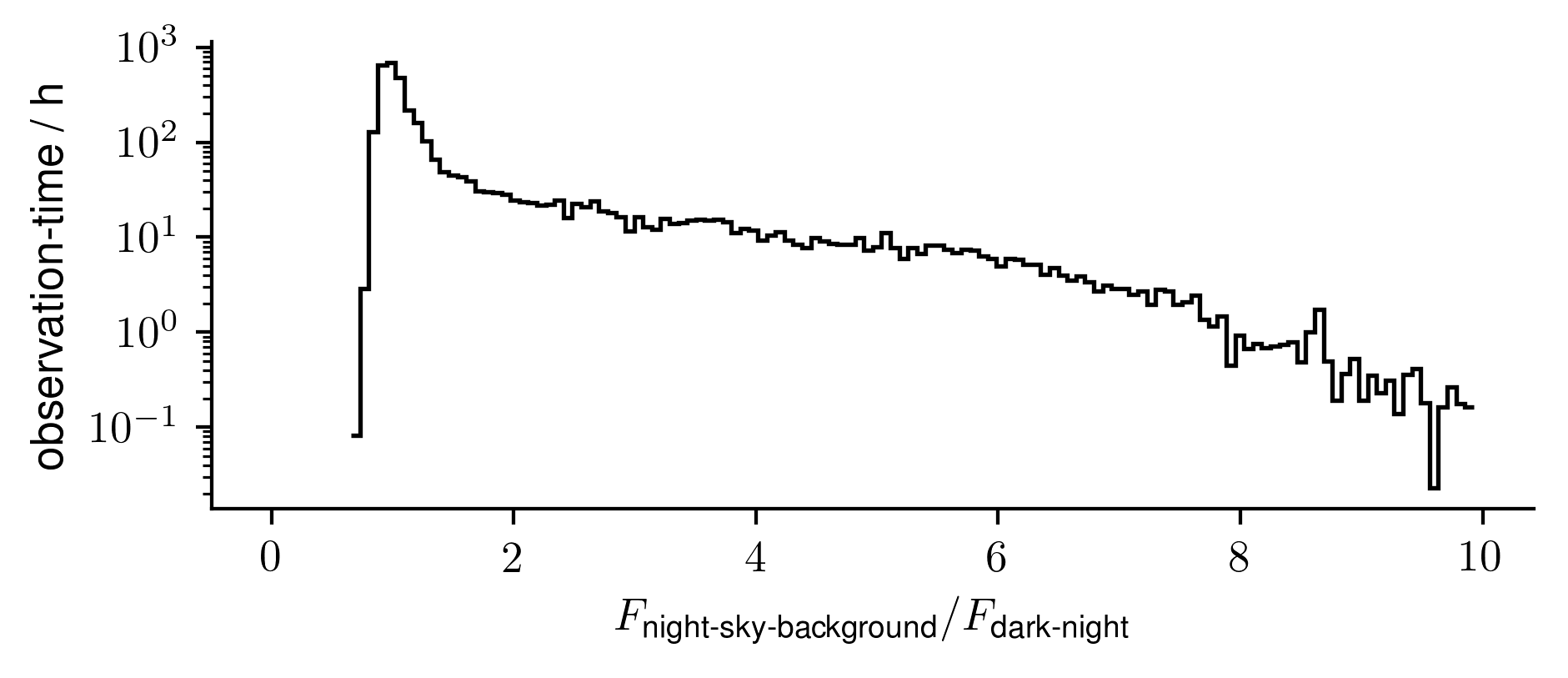}
                    \caption{Observation time for different night-sky-background rates (dark night = $35 \times 10^6$ photons $s^{-1} pixel^{-1}$) \cite{mueller2019cherenkov}}
                    \label{fig:observationTime}
                \end{center}
            \end{figure}

            Although for night-sky-background rate of $70 \times 10^6$ photons $s^{-1} pixel^{-1}$ the precision is $\sim 70\%$, we can see from \ref{fig:observationTime}, then most of the observations are done with less night-sky-background, thus having higher precision.

    \section{Extraction of muon ring properties}

\begin{figure}[H]
                \begin{center}
                    \includegraphics[width=\linewidth]{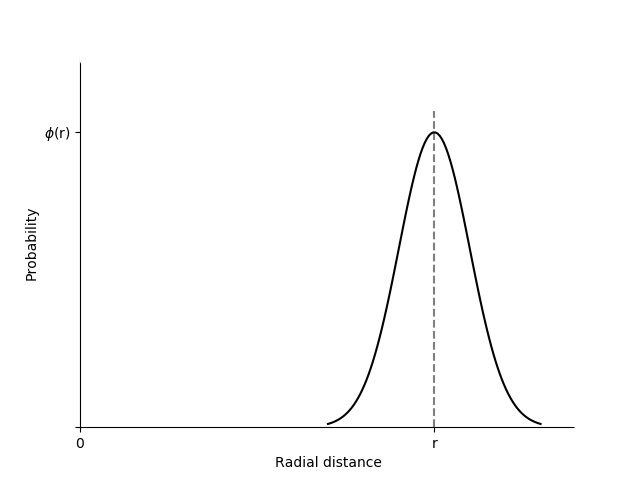}
                    \caption{Sharp ring (dashed grey line) is blurred (solid black line) due to pixellation of the camera and misaligned reflector facets}
                    \label{fig:fuzziness}
                \end{center}
            \end{figure}

        The goal of extracting the muon ring features (extraction for short) is to get the radius and the center position of the muon ring. Cherenkov photons are expected to form a sharp ring on the sensor plane. Due to camera's pixellation and misaligned reflector facets the ring image is blurred (see figure \ref{fig:fuzziness}). To describe the spread, a fuzziness parameter is used.

        Accurate reconstruction of the ring parameters is highly preferable to avoid large uncertainties in the fuzziness parameter. In contrast to detection, the choice for extraction methods is much wider Methods of extraction include for example a simple Circle Model and Hough Transformation.

    \section{Circle Model (RM)}

        The Circle model used in this thesis is taken directly from skimage.measure package (\cite{skimage}). It estimates 2D circles using total least squares method. The functional model for the circle is 

        \begin{equation}
        r^2 = (x - x_c)^2 + (y - y_c)^2
        \end{equation}

        This model minimizes the squared distances from all the points to the circle:

        \begin{equation}
        min \{ \sum(r - \sqrt[]{(x - x_c)^2 + (y - y_c)^2}) \}
        \end{equation}
        For this model a minimum of 3 photons is required. (See examples in \cite{CircleModel})

    \section{Hough transform(HT)}
        Hough transform (\cite{duda1972use}) is a technique used in computer vision to identify objects from a background. This technique finds imperfect instances of objects that lie within a certain class. The best candidate is chosen via a integrating procedure that is carried out in a parameter space (called Hough accumulator or Hough space).

\begin{figure}[H]
                \begin{center}
                    \includegraphics[width=\linewidth]{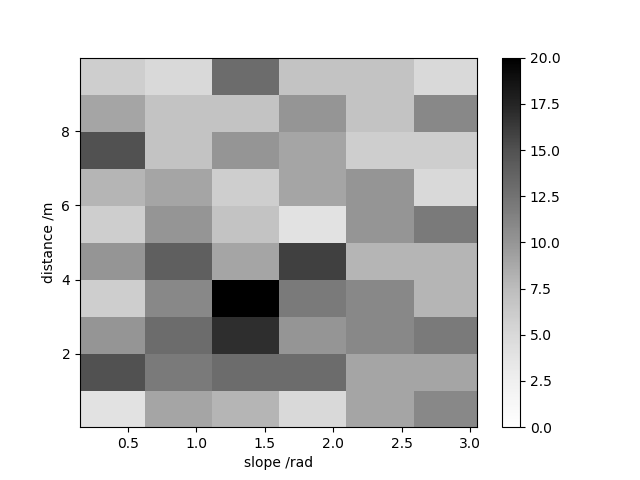}
                    \caption{Example of 2D parameter space (Hough accumulator) for a line. Darkest bin corresponds to best parameters for a line}
                    \label{fig:param_space2d}
                \end{center}
            \end{figure}

        Dimension of parameter space depends on the amount of variables needed to describe a certain object. For example a line can be described by 2 parameters: slope and distance from the origin, thus the parameter space is two dimensional. Visually it would be represented like a 2D-plane (see figure \ref{fig:param_space2d})

        Collecting all votes from the parameter space we obtain the local maxima, whose coordinates in this accumulator correspond to the best fit of the sought-after object (be it a line, circle or something else).

        \subsection{Hough transform for circle}
            Since in this thesis the goal is to infer true optical pointspread function using muon rings, then in order to reconstruct the rings we need Hough transform for recognizing rings. As a first attempt the accumulator was filled with votes that come from astropy Ring2D fit (\cite{astropy}. Unfortunately to achieve good enough accuracy the grid of the accumulator needed to be very fine. Consequently this increased also the computational time needed. 

\begin{figure}[H]
                    \begin{center}
                        \includegraphics[width=0.6\linewidth]{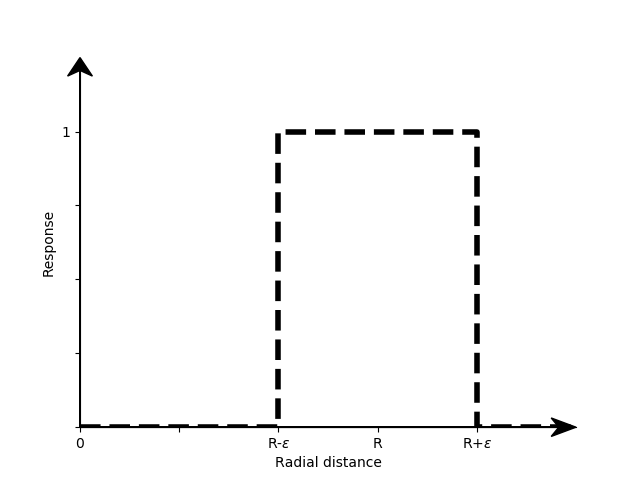}
                        \caption{Heaviside weighing function}
                        \label{fig:heavyside}
                    \end{center}
                \end{figure}

\begin{figure}[H]
                    \begin{center}
                        \includegraphics[width=0.6\linewidth]{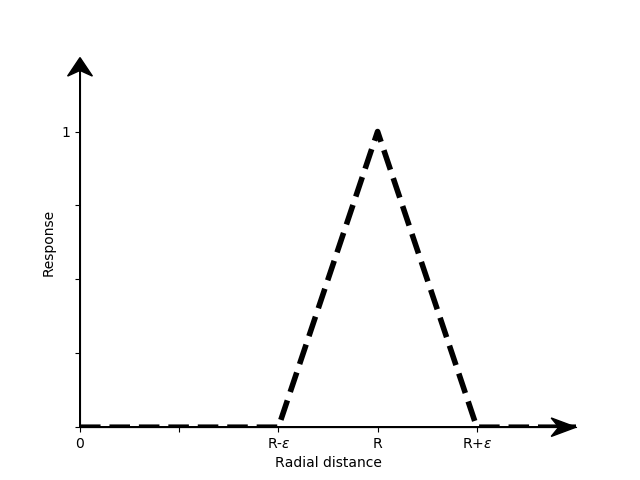}
                        \caption{Triangular weighing function}
                        \label{fig:triangular}
                    \end{center}
                \end{figure}

            To overcome this problem a Hough transform script was written, that had some extra features: instead of the Heaviside weighing function (\cite{heaviside}) (see figure \ref{fig:heavyside}) triangular weighing function was used (see figure \ref{fig:triangular}), thus avoiding same value bin-plateaus in the Hough accumulator, thereby making it possible to choose one best parametrization for the ring. This increases the accuracy when deciding which of the accumulator bins had the most votes.

            Furthermore our custom Hough transform is iterative, so in every interation-step the area to be scanned gets smaller but with the same bin-count. Thus computing time is reduced and higher accuracy is reached. 

            Finally, an initial guess can be given to the transformation together with the total uncertainty, thus decreasing the amount of parameter space to be scanned by the Hough transformation. In case when no initial guess is available, one should adjust the uncertainty according to the amount of area to be scanned. (See more details in \cite{circlehough})

    \section{Used combinations of detection and extraction and their properties}
        In the following one can see the results of 300 000 muons that were simulated and different detection and extraction method combinations were tested.

        \subsection{DB-RM}
        The most straight-forward method to test was density clustering with a simple ring model. As the name suggests then in order to detect Cherenkov photons from the night-sky-background photons, this method uses density clustering (DBSCAN algorithm to be more exact). Found Cherenkov photons are then used to extract muon rings and their features. The first estimate is done by using a regular CircleModel from skimage which is wrapped with RanSaC (Random Sampling Consensus) (\cite{fischler1981random}) to get rid of the outliers.

\begin{figure}[H]
                \begin{center}
                    \includegraphics[width=0.7\textwidth]{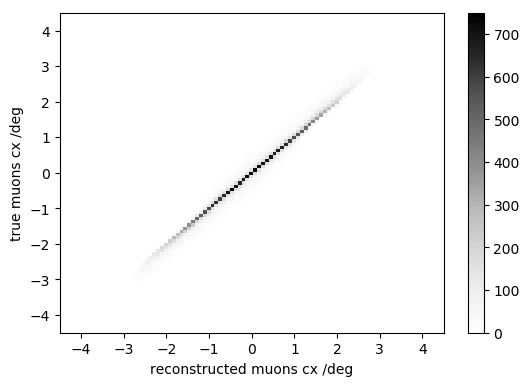}
                    \caption{Confusion matrix for cx with DB-RM method}
                    \label{fig:ex_r_cx}
                \end{center}
            \end{figure}

\begin{figure}[H]
                \begin{center}
                    \includegraphics[width=0.7\textwidth]{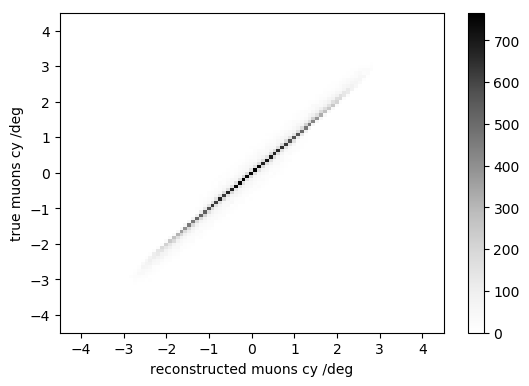}
                    \caption{Confusion matrix for cy with DB-RM method}
                    \label{fig:ex_r_cy}
                \end{center}
            \end{figure}

        As one can see from figure \ref{fig:ex_r_cx} and figure \ref{fig:ex_r_cy} at first glance it would seem that the center positions of the muon ring are reconstructed fairly well. Unfortunately this cannot be said about the ring radius. When one looks closer at the scale, then it becomes clear that ring center positions are reconstructed with an error of rougly three quarters of a degree. This error causes also the ring radius to be reconstructed very inaccurately and one can even see a major bias - this method underestimates the ring radius very often.

\begin{figure}[H]
                \begin{center}
                    \includegraphics[width=0.7\textwidth]{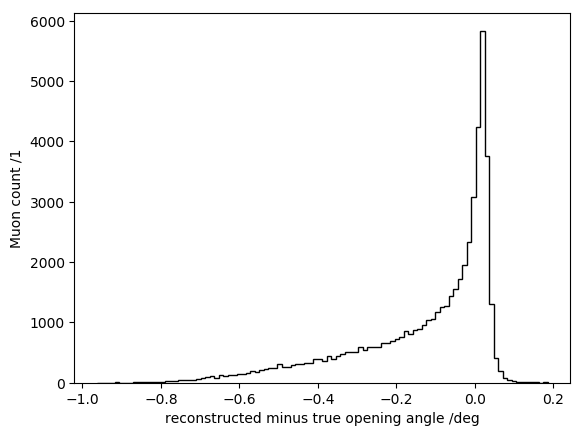}
                    \caption{Reconstructed opening angle minus true opening angle}
                    \label{fig:ringM_oa}
                \end{center}
            \end{figure}

        When subtracting true opening angle from reconstructed one, one sees how inaccurately the radius is reconstructed (see figure \ref{fig:ringM_oa}). The standard deviation of the distribution for the difference of reconstructed opening angle and true opening angle is 0.183 degrees.

\begin{figure}[H]
                \begin{center}
                    \includegraphics[width=0.7\textwidth]{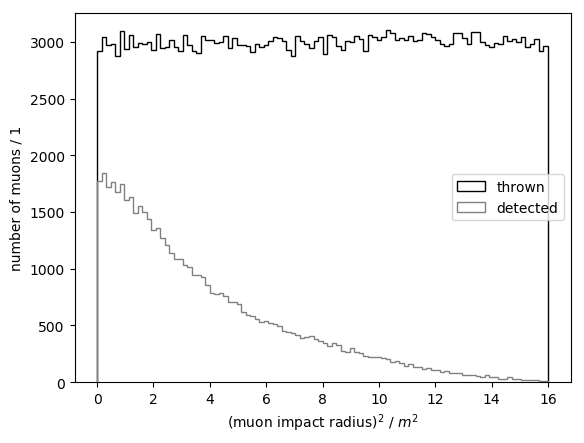}
                    \caption{Impact radius squared with DB-RM method}
                    \label{fig:ex_r_r}
                \end{center}
            \end{figure}

\begin{figure}[H]
                \begin{center}
                    \includegraphics[width=0.7\textwidth]{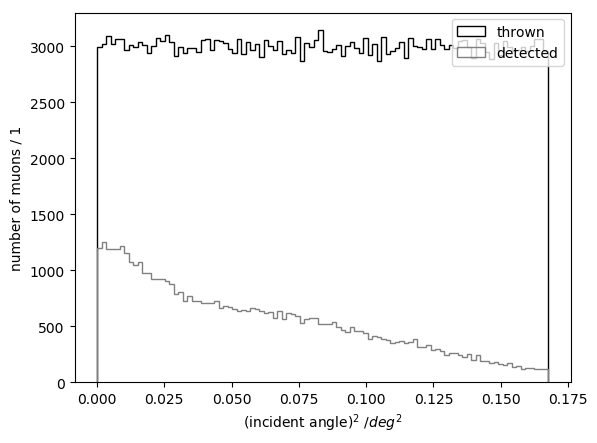}
                    \caption{Ring center coordinates squared with DB-RM method}
                    \label{fig:ex_r_cxcy}
                \end{center}
            \end{figure}

        On figure \ref{fig:ex_r_r} we can see that indeed the simulated muons support position on apperture plane is distributed uniformly over the disk with the given aperture radius (4 m). As one would expect then the muons which hit the aperture plane closer to the center were detected more efficiently.
        On figure \ref{fig:ex_r_cxcy} we see, that the simulated muons had directions uniformly distributed and smaller inclination angles were more likely to be detected.

            \begin{figure}[H]
                \begin{center}
                    \includegraphics[width=0.7\textwidth]{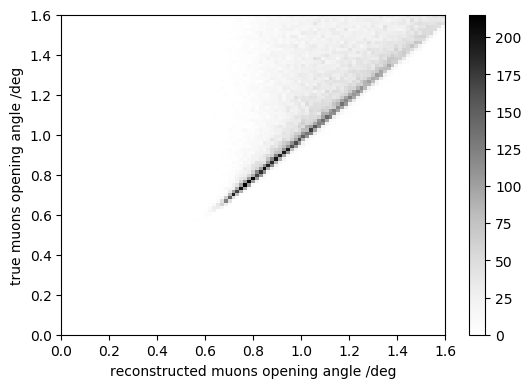}
                    \caption{Confusion matrix for opening angle with DB-RM method}
                    \label{fig:ex_r_oa}
                \end{center}
            \end{figure}

        \subsection{DB-RM-HT}

        The custom Hough transform script uses initial guess for improving the accuracy of the reconstruction. As a byproduct it reduces computing time. The initial guess for ring parameters is taken from the DB-RM results. To increase accuracy even more, one performs multiple Hough transforms (iterative Hough, decreasing bin sizes with every step), taking the results from the previous iteration to be the initial guess and thus reducing the amount of Hough space to be scanned each iteration.

\begin{figure}[H]
            \begin{center}
                \includegraphics[width=\textwidth]{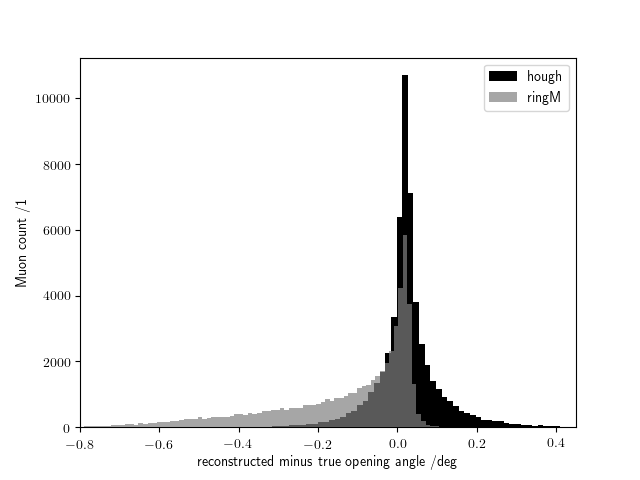}
                \caption{Opening angle of the Cherenkov cone}
                \label{fig:hough_oa}
            \end{center}
        \end{figure}

\begin{figure}[H]
            \begin{center}
                \includegraphics[width=0.7\textwidth]{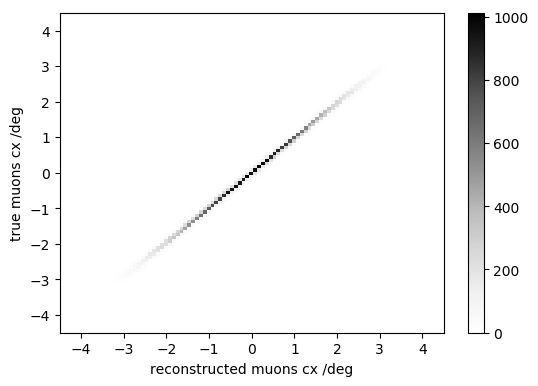}
                \caption{Confusion matrix for cx with DB-RM-HT method}
                \label{fig:circleHough_cx}
            \end{center}
        \end{figure}

\begin{figure}[H]
            \begin{center}
                \includegraphics[width=0.7\textwidth]{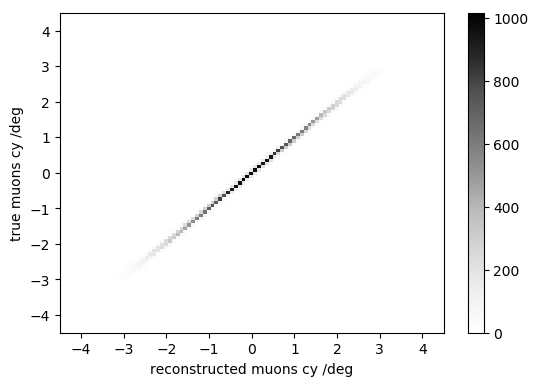}
                \caption{Confusion matrix for cy with DB-RM-HT method}
                \label{fig:circleHough_cy}
            \end{center}
        \end{figure}

        As one can see from figures \ref{fig:circleHough_cx} and \ref{fig:circleHough_cy} it is clear that the extracted ring center position is very accurate. Furthermore, when one compares reconstructed ring radius with Hough transform and the one from Circle Model (DB-RM) on figure \ref{fig:hough_oa} it is obvious that Hough transform version is superior, having much narrower and symmetric distribution.

        In addition in order to reach this accuracy without iterative Hough, one needs $\sim100$ bins for all of the 3 variable, thus $10^6$ bins needed to be checked. With iterative Hough only $\sim8000$ bins needs to be checked, thus the computing time reduced by a factor of 125 at least.

        \begin{figure}[H]
            \begin{center}
                \includegraphics[width=0.7\textwidth]{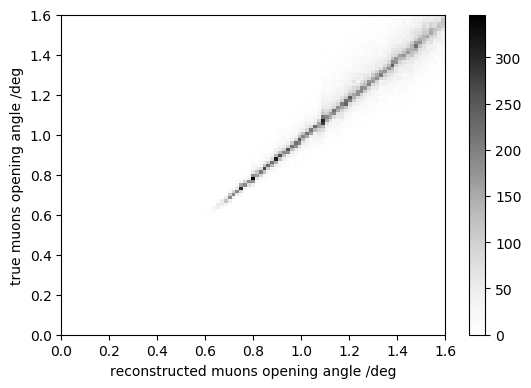}
                \caption{Confusion matrix for opening angle with DB-RM-HT method}
                \label{fig:circleHough_oa}
            \end{center}
        \end{figure}

        \begin{figure}[H]
            \begin{center}
                \includegraphics[width=0.7\textwidth]{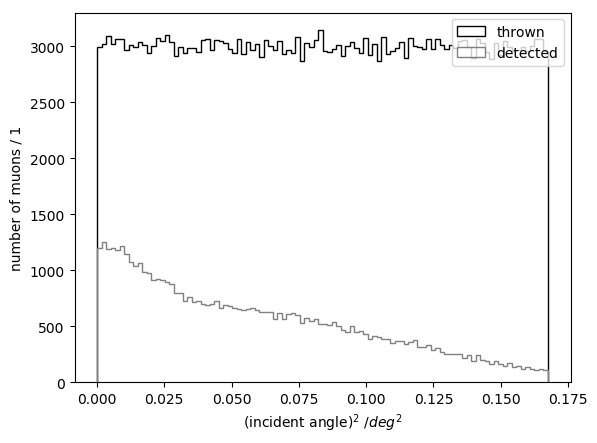}
                \caption{Ring center coordinates squared with DB-RM-HT method}
                \label{fig:circleHough_cxcy}
            \end{center}
        \end{figure}

\begin{figure}[H]
            \begin{center}
                \includegraphics[width=0.7\textwidth]{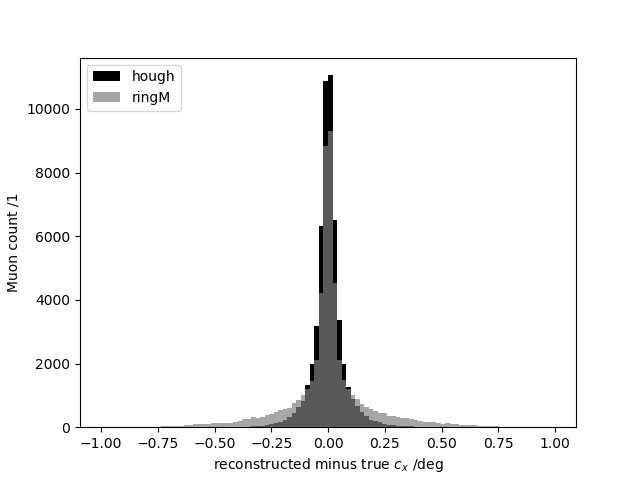}
                \caption{Reconstructed cx minus true cx}
                \label{fig:ringM_hough_cx}
            \end{center}
        \end{figure}

        Comparison of DB-RM and DM-RM-HT ring center reconstructions are found on figures \ref{fig:ringM_hough_cx} and \ref{fig:ringM_hough_cx}. The standard deviations of the distributions are in table \ref{tab:cxcy-gauss}.

        \begin{figure}[H]
            \begin{center}
                \includegraphics[width=0.7\textwidth]{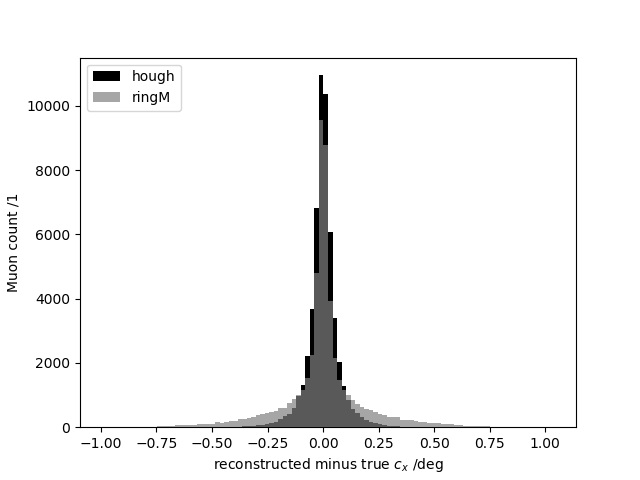}
                \caption{Reconstructed cy minus true cy}
                \label{fig:ringM_hough_cy}
            \end{center}
        \end{figure}

        \begin{table}[H]
          \begin{center}
            \begin{tabular}{l l l}
              Method & $\sigma(c_x) (degrees)$ & $\sigma(c_y) (degrees)$\\
              \hline
              DB-RM-HT & 0.0712 & 0.0703\\
              DB-RM & 0.183 & 0.180\\
            \end{tabular}
            \caption{Standard deviation of the difference of true and reconstructed ring center}
            \label{tab:cxcy-gauss}
          \end{center}
        \end{table}

    \section{Conclusions of the ring reconstruction methods}

        In table \ref{tab:gauss} are listed the standard deviations of the difference of true and reconstructed opening angles. It is evident, that DB-RM-HT method is superior to other tried methods, thereby reconstructing rings $\sim 2.5$ times more accurately.

\definecolor{light-gray}{gray}{0.80}
        \begin{table}[H]
          \begin{center}
            \begin{tabular}{l c}
              Method & \textbf{$\sigma (degrees)$}\\
              \hline
              \textcolor{light-gray}{DB-KC-SD \tablefootnote{see appendix \ref{KC}}} & \textcolor{light-gray}{0.0202}\\
              DB-RM-HT & 0.0730\\
              DB-RM & 0.183\\
              \textcolor{light-gray}{DB-RM-SD \tablefootnote{see appendix \ref{SD}}} & \textcolor{light-gray}{0.187}
            \end{tabular}
            \caption{Standard deviation of the difference of true and reconstructed opening angle}
            \label{tab:gauss}
          \end{center}
        \end{table}

    \section{Classification of the event}

        For classification of the events the following cuts are applied after Cherenkov photon detection and ring feature extraction:

        \begin{table}[H]
          \begin{center}
            \begin{tabular}{l r}
              \textbf{Parameter} & \textbf{cut}\\
              \hline
              number-of-photons-min & 3 \\
              muon-ring-radius-min & 0.45 deg\\
              muon-ring-radius-max & 1.6 deg\\
              muon-ring-overlapp-with-field-of-view-min& 20\%\\
              arrival-time-stddev-max & 5e-9 s\\
              initial-circle-model-photon-ratio-min & 0.6 \\
              visible-ring-circumfance-min & 1.5 deg\\
              off-density & 0 \\
              on-off-ratio-min & 3.5 \\
              density-circle-model-inner-ratio-max & 0.25 \\
              evenly-population & True \\
            \end{tabular}
            \caption{Current cut parameters to find muons}
            \label{tab:cut-parameters}
          \end{center}
        \end{table}

        where:
        \begin{itemize}
            \item \textbf{number-of-photons-min}: Minimum number of photons required to perform Circle Model.
            \item \textbf{muon-ring-radius-min}: Minimum muon ring radius for it to be classified as muon ring.
            \item \textbf{muon-ring-radius-max}: Maximum muon ring radius for it to be classified as muon ring.
            \item \textbf{muon-ring-overlapp-with-field-of-view-min}: Minimum muon ring overlap with the field of view (fraction).
            \item \textbf{arrival-time-stddev-max}: Maximum arrival time standard deviation.
            \item \textbf{initial-circle-model-photon-ratio-min}: Minimum fraction of inlier photons divided by all photons.
            \item \textbf{visible-ring-circumfance-min}: Minimum visible ring circumference in degrees.
            \item \textbf{off-density}: average density of photons offset from the ring circumference.
            \item \textbf{on-off-ratio-min}: Minimal fraction of on-density (points on ring circle) divided by off-density
            \item \textbf{density-circle-model-inner-ratio-max}: Maximum fraction of number of photons off ring inside circle divided by number of all photons.
            \item \textbf{evenly-population}: Check whether ring is populated evenly. Threshold for evenly-population to be true is if the standard deviation is less than 0.8
        \end{itemize}

        With these cuts, the performance of classifying events to be muons had the following performance:

        \begin{table}[H]
          \begin{center}
            \begin{tabular}{l l}
              Precision & 99.3\%\\
              Sensitivity & 67.2\%\\
            \end{tabular}
            \label{tab:performance}
          \end{center}
        \end{table}

        \subsection{Acceptance vs opening angle}
\begin{figure}[H]
                \begin{center}
                    \includegraphics[width=0.8\textwidth]{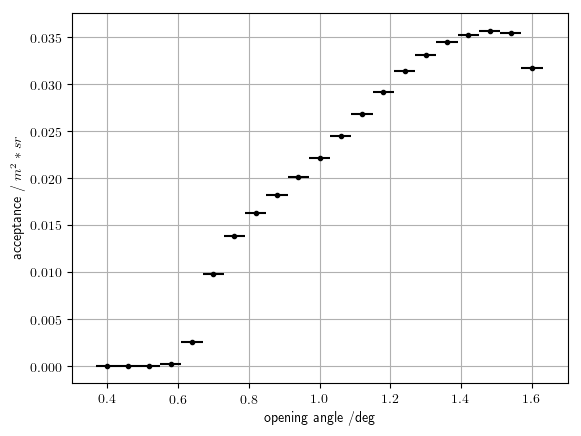}
                    \caption{Acceptance vs opening angle}
                    \label{fig:effective_area_vs_oa}
                \end{center}
            \end{figure}

            For the best found method, DB-RM-HT, also muon acceptance dependent on the opening angle of the Cherenkov cone was found by simulating muons with a set opening angle. As one can see from figure \ref{fig:effective_area_vs_oa} then the effective area peaks near $\sim 1.2$ degrees, which is also the maximum angle under which muons in at the height of the Telescope can emit Cherenkov emission.

        \subsection{Acceptance vs point spread function}

\begin{figure}[H]
                \begin{center}
                    \includegraphics[width=0.8\textwidth]{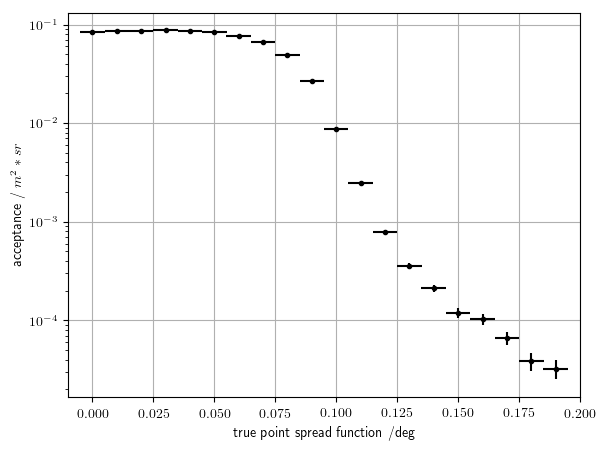}
                    \caption{Acceptance vs point spread function}
                    \label{fig:effective_area_vs_psf}
                \end{center}
            \end{figure}

            Also muon acceptance dependence was studied for different point spread functions. As one can see from figure \ref{fig:effective_area_vs_psf} then the acceptance reduces when muon ring image gets fuzzier.

    \section{PSF vs fuzziness}\label{sec:psf_fuzz}

        Since the goal of the thesis is to measure PSF using muon rings, then we need to relate muon ring fuzziness parameter to the actual PSF. Since perfect muon rings should be sharp, then the task is easily achievable by simulating muons with different point spread functions and relating them to the calculated fuzziness. In total 300 000 muons were simulated for every different point spread function between 0 and 0.20 degrees PSF with 20 steps.

        We expect the fuzz vs. PSF plot to have a 3 regions that can be approximated with lines with different slopes. This is expected because in the region with very small point spread function it will be saturated due to the pixellation (\cite{kamgar1989quantization}). The theoretical limit for the point spread function caused by pixellation would be thus $\sigma_{pixel} = \frac{1}{\sqrt{12}} \times \theta = 0.0289$. The second region is expected to be linear because of the relation between fuzziness PSF should be linear. Third region starts roughly when the PSF is the size of the pixel and thus unexpected behaviour is expected. Thus it makes sense to use a third degree polynomial fit, because this is the lowest order polynomial that has two curvatures. Because we want to have still some info about saturated parts, we will keep some of it and fit a polynomial of third degree.

        \subsection{Standard deviation as fuzziness parameter} \label{sec:stdev}

            This fuzziness parameter is the standard deviation of the distances of all the photons from the ringline.

\begin{figure}[H]
                    \begin{center}
                        \includegraphics[width=0.7\linewidth]{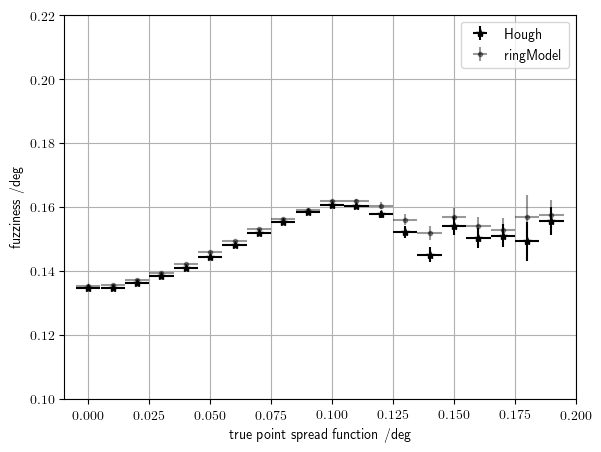}
                        \caption{Fuzziness vs point spread function}
                        \label{fig:nsb35e6a}
                    \end{center}
                \end{figure}

            On figure \ref{fig:nsb35e6a} one can see a clear fuzziness dependence on point spread function, although there is a slight saturation with very low point spread function caused by the pixellation of the camera.

            Fit for standard deviation as fuzziness parameter:

            \begin{equation}\label{eq:stdev_fuzz}
                -53.5 x^3 + 9.14x^2 - 1.32\times10^{-1}x + 1.35\times10^{-1}
            \end{equation}

            \begin{figure}[H]
                \begin{center}
                    \includegraphics[width=0.7\linewidth]{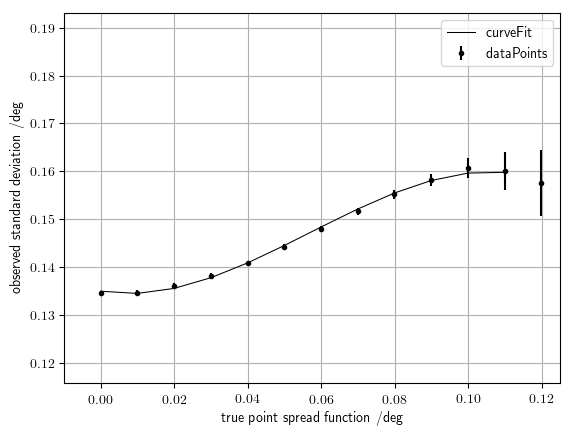}
                    \caption{Standard deviation as fuzziness parameter}
                    \label{fig:curve_fitting_stdev}
                \end{center}
            \end{figure}

        \subsection{Hough response as fuzziness parameter}

            The fuzziness parameter called `Hough response' or just `response' is the normed value of the highest value bin in the Hough accumulator (Hough space). It is calculated as:

            \begin{equation}
                \frac{highest\_value\_bin}{all\_photons\_count} * 100
            \end{equation}

            Thus for the sharpest ring the response would be 100

            Fit for response as the fuzziness parameter:

            \begin{equation}\label{eq:response_fuzz}
                2.77\times10^4 x^3 -2.76\times10^3x^2 - 2.86\times10^2x + 51.3
            \end{equation}

            \begin{figure}[H]
                \begin{center}
                    \includegraphics[width=0.7\linewidth]{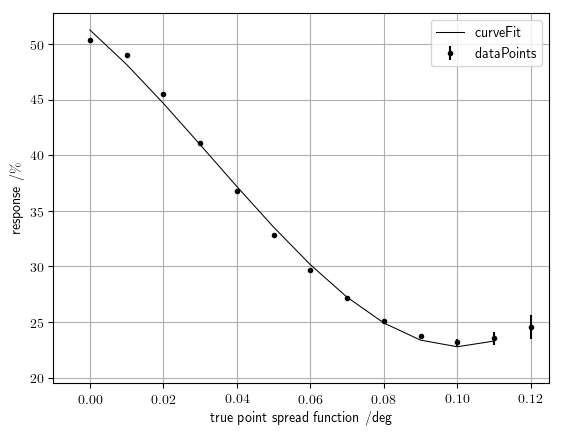}
                    \caption{Response as the fuzziness parameter}
                    \label{fig:curve_fitting_response}
                \end{center}
            \end{figure}

            \begin{figure}[H]
                \begin{center}
                    \includegraphics[width=0.7\linewidth]{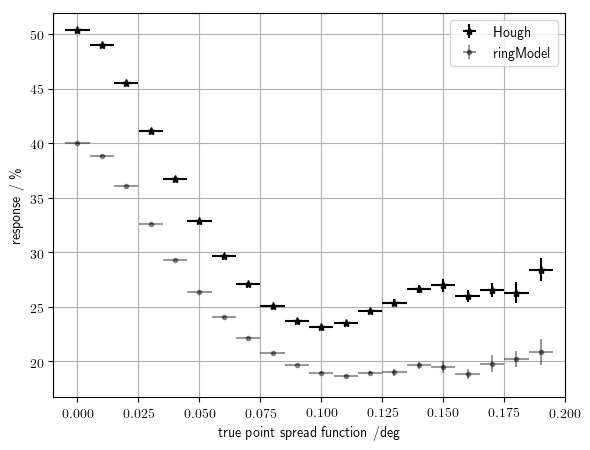}
                    \caption{Response as the fuzziness parameter}
                    \label{fig:respose_psf}
                \end{center}
            \end{figure}

            Comparing DB-RM (ringModel) and DB-RM-HT (Hough) on figure \ref{fig:respose_psf} we have an indication that Hough transform would outperform ringModel by having steeper slope in the correlation of fuzziness vs. point spread function.

    \section{Comparison of fuzz parameters}

\begin{figure}[H]
            \begin{center}
                \includegraphics[width=0.8\linewidth]{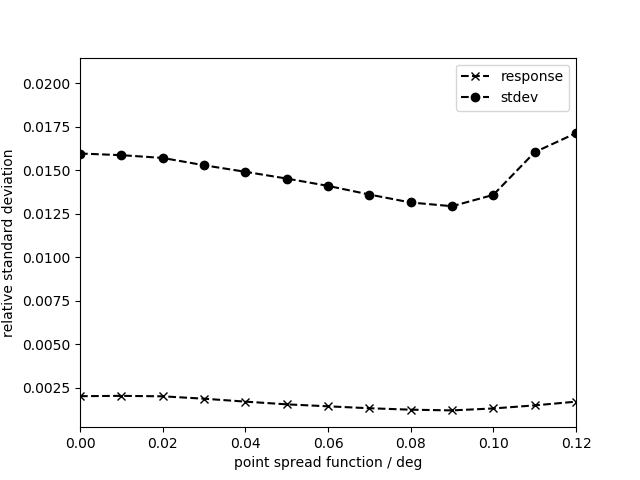}
                \caption{Relative standard error for fuzziness parameter value vs psf}
                \label{fig:relative_fuzz_error}
            \end{center}
        \end{figure}

        As one can see from figure \ref{fig:relative_fuzz_error} an important advantage of the fuzziness parameter 'response' over 'stdev' (see subsection \ref{sec:stdev}) is that the relative standard deviation is smaller manyfold, thus one would expect much more accurate results when reconstructiong PSF, which we will see in the next chapter.

\newpage
\chapter{Real observations}

    \section{Fuzziness parameters and PSF vs time}
        As we saw in chapter \ref{chap:investigation} that Hough transform (DB-RM-HT) outperforms ringM (DB-RM) in reconstructing ring parameters. In the following only results using Hough transform as the extraction method are shown. To arrive at our goal to see how point spread function varies in time we have to go through four steps:

            \begin{itemize}
                \item Simulate event with different point spread functions to get the relation between our chosen fuzziness parameter and true PSF (See section \ref{sec:psf_fuzz})
                \item Fit a curve to infer PSF (see also section \ref{sec:psf_fuzz})
                \item Find fuzziness parameter over time 
                \item Convert fuzziness to PSF
            \end{itemize}

        \subsection{Standard deviation as the fuzziness parameter}

\begin{landscape}
                \begin{figure}[H]
                    \includegraphics[width=\linewidth]{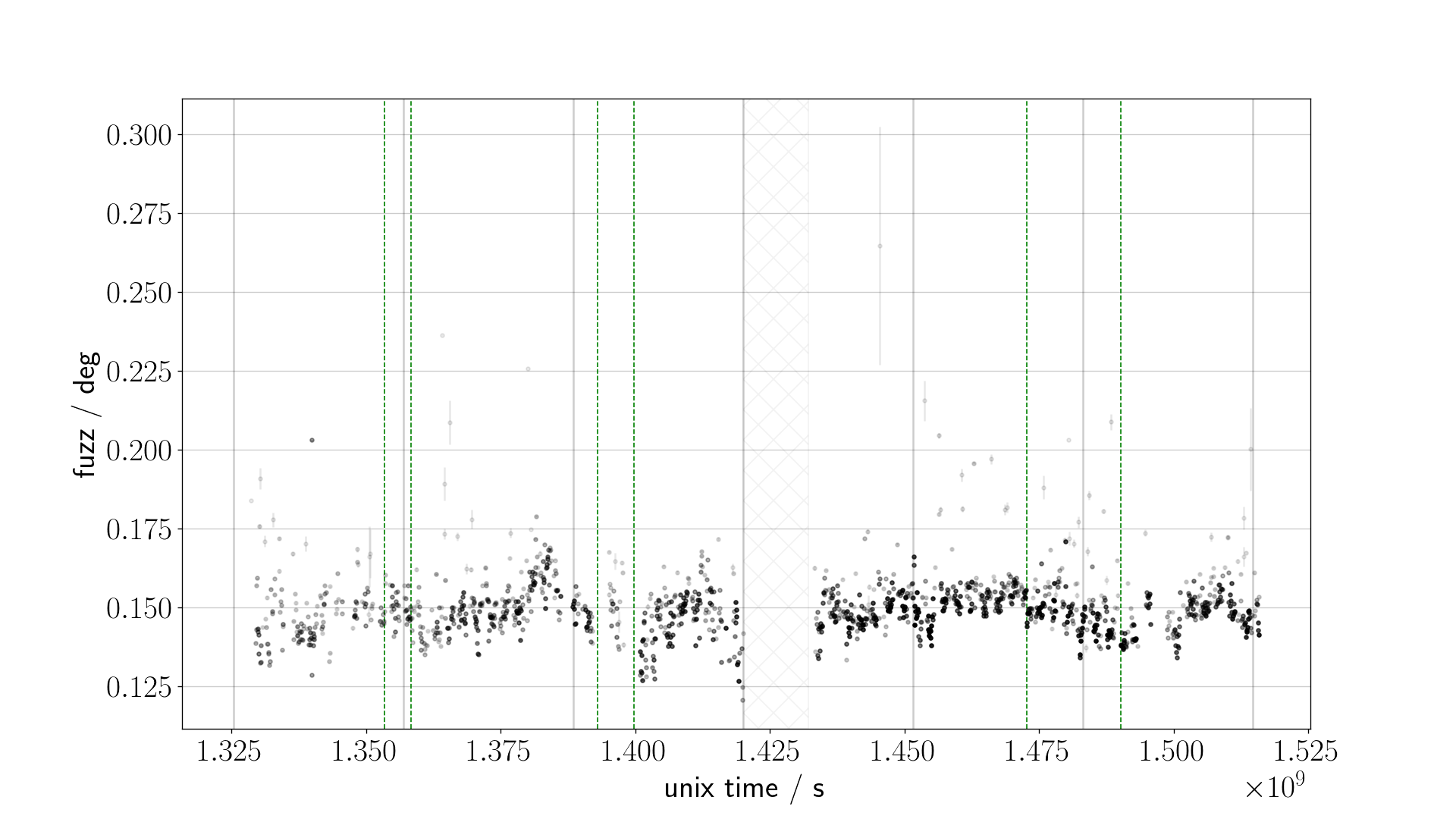}
                    \caption{Standard deviation as the fuzziness parameter over time}
                    \label{fig:hough_fuzz_stdev}
                \end{figure}
            \end{landscape}

            As the first step we calculate the fuzziness parameter (currently standard deviation of the photon from the reconstructed ring) for all the observations. This gives us figure \ref{fig:hough_fuzz_stdev}. The dimmer a datapoint is, the less significant the date is, because the less muons were detected. The size of the errorbar around the datapoint is indicates how big was the standard error of the fuzziness parameter at that date. Already from this plot we can see rougly how true point spread function would behave. The green lines on the plot show timestamps where mirror alignments were done. Furthermore there is a hatched area on the figure, which shows when the data taken was not reliable due to bad DRS (Domino Ring Sampler) calibration.

        \subsection{Hough response as the fuzziness parameter}

\begin{landscape}
                \begin{figure}[H]
                    \includegraphics[width=\linewidth]{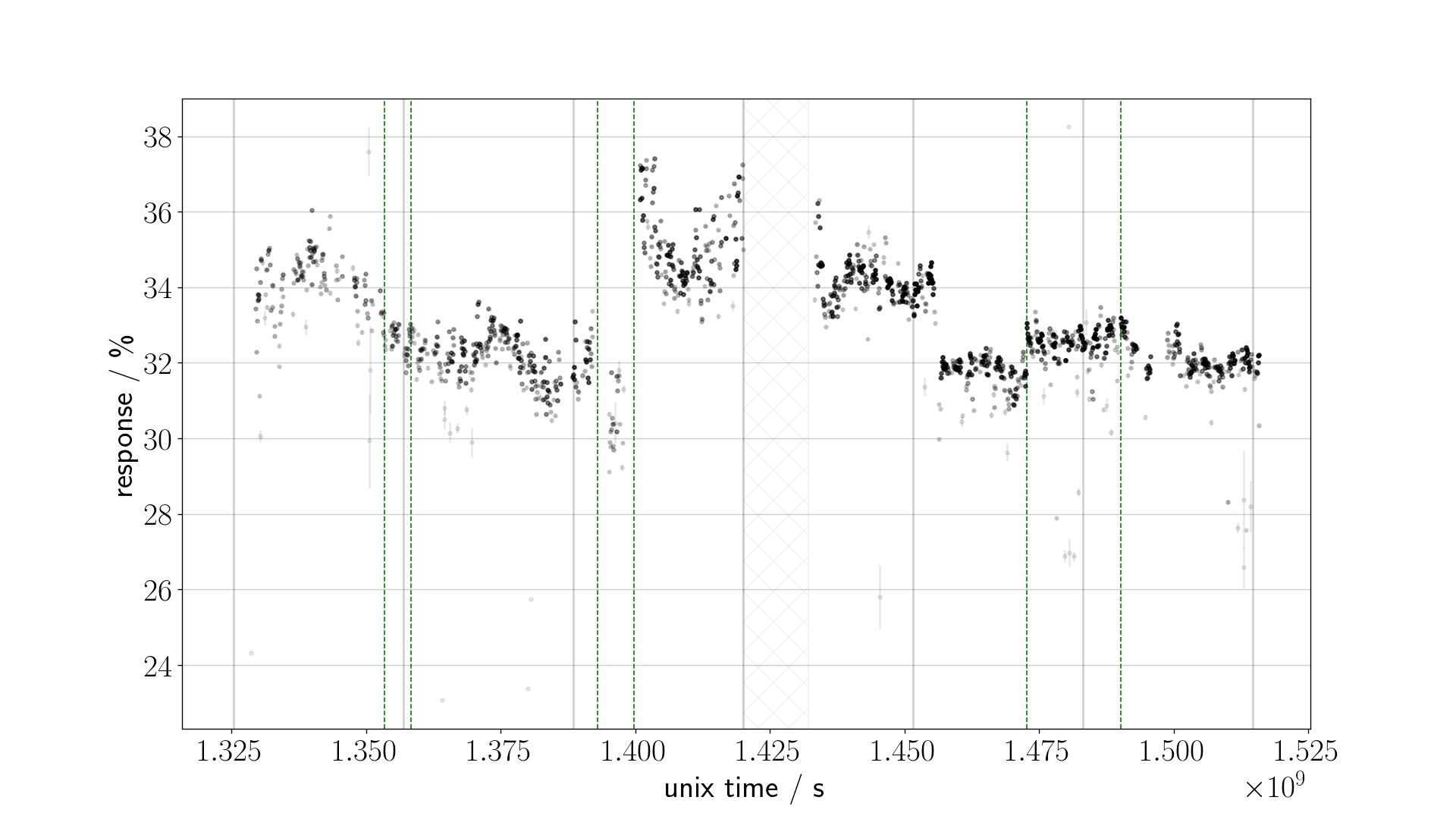}
                    \caption{Response as the fuzziness parameter over time}
                    \label{fig:hough_fuzz_response}
                \end{figure}
            \end{landscape}

\begin{landscape}
                \begin{figure}[H]
                    \includegraphics[width=\linewidth]{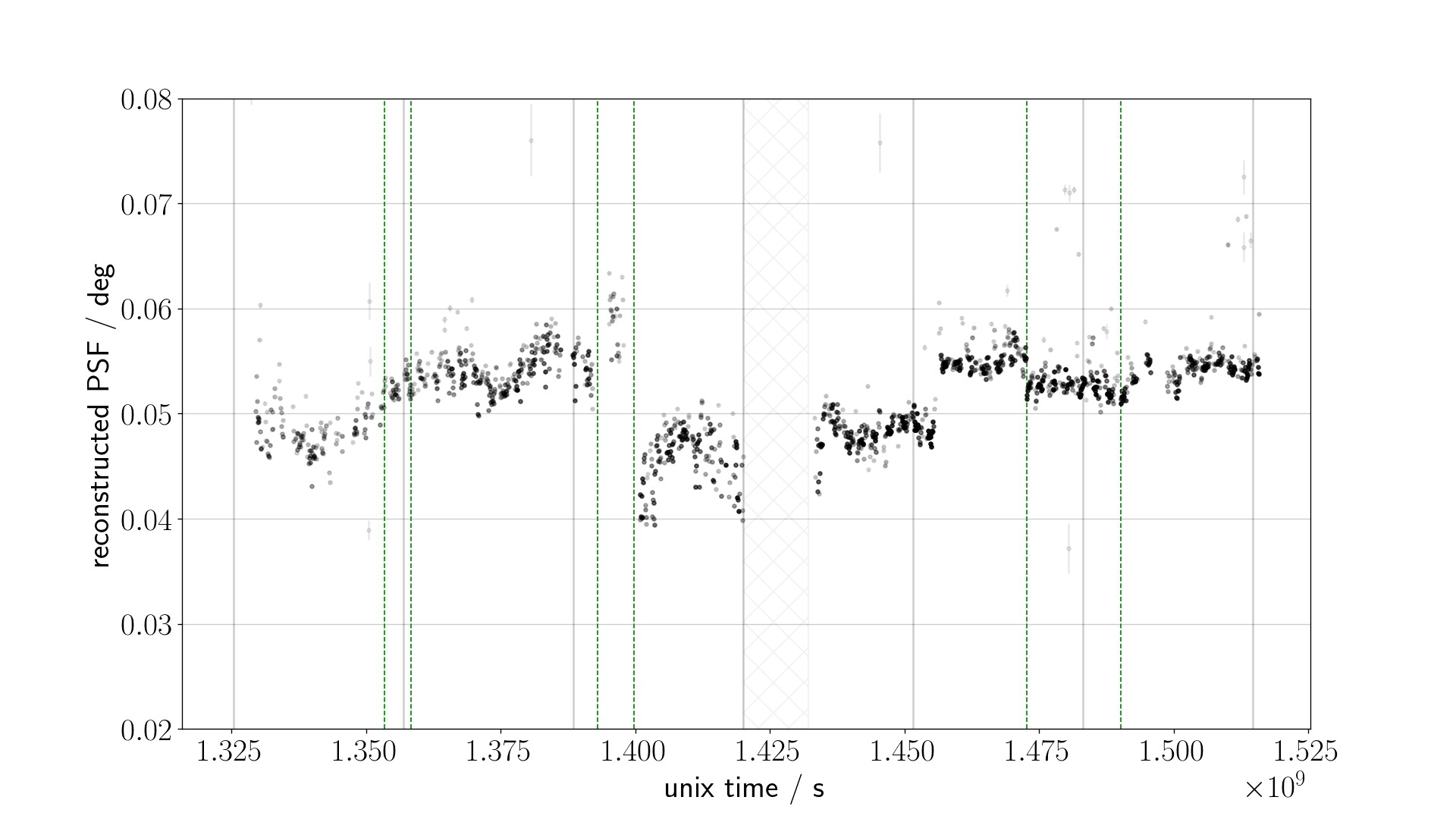}
                    \caption{Reconstructed PSF using response as the fuzziness parameter vs time}
                    \label{fig:hough_psf_response}
                \end{figure}
            \end{landscape}

            As seen on the figure \ref{fig:hough_fuzz_response}, it is clear that this fuzziness parameter distinguishes different point spread functions very well, thus changes in PSF should be visible. One should not forget, that the higher the response, the smaller the point spread function. Furthermore, as was shown on figure \ref{fig:relative_fuzz_error}, then we would expect not much noise in the PSF vs. time figure. Those assumptions turn out to be true, as one can see from figure \ref{fig:hough_psf_response}.

\begin{figure}[H]
                \includegraphics[width=\linewidth]{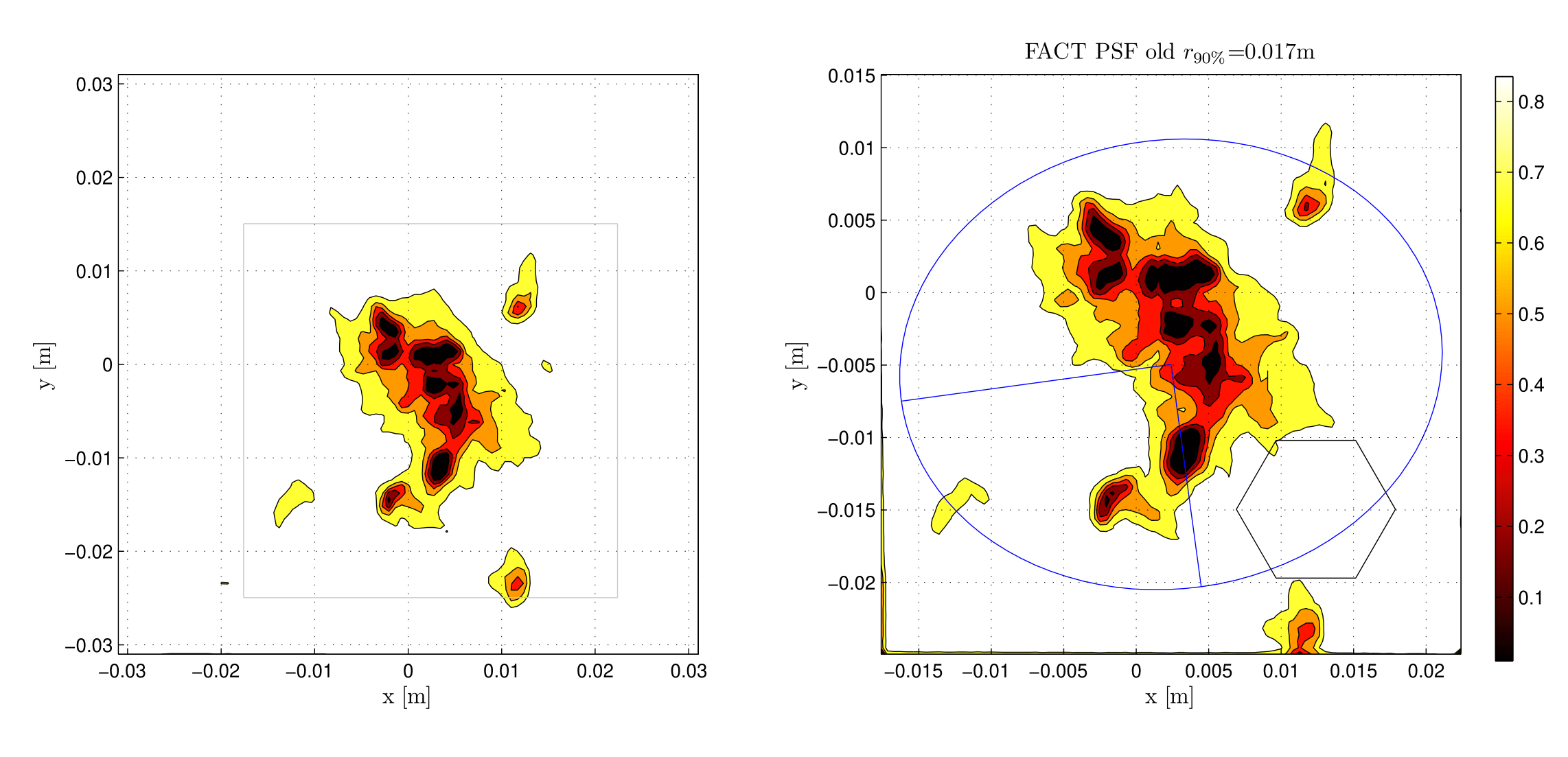}
                \caption{PSF before 10.05.2015 alignment \cite{muller2014clear}}
                \label{fig:before}
            \end{figure}

\begin{figure}[H]
                \includegraphics[width=\linewidth]{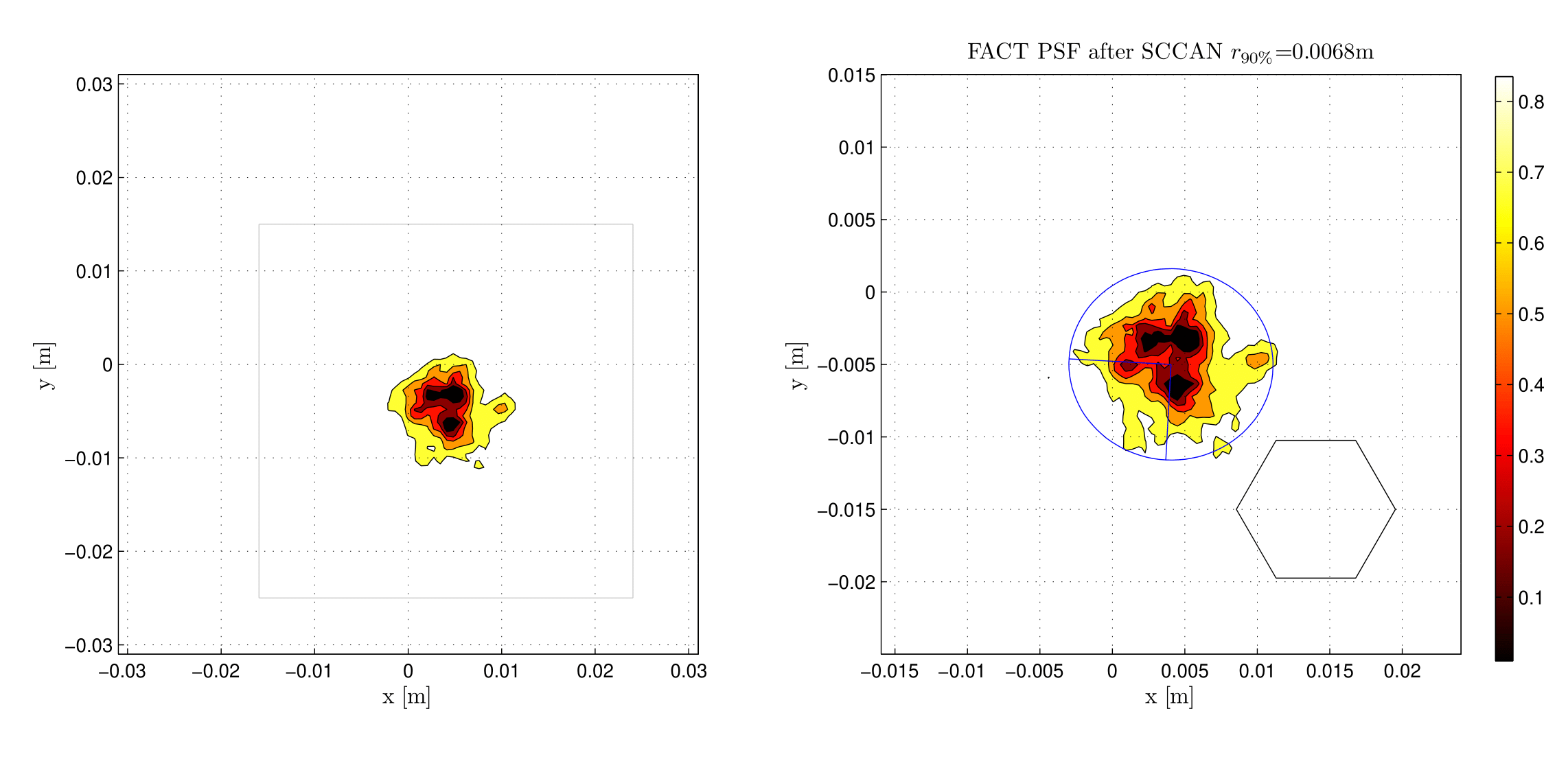}
                \caption{PSF after 10.05.2015 alignment \cite{muller2014clear}}
                \label{fig:after}
            \end{figure}

            Green dashed lines on figures \ref{fig:hough_fuzz_response}, \ref{fig:hough_psf_response} indicate times when a mirror alignment was done. We can see that after the most thorough mirror realignment at 14th of May 2015 (fourth green dashed line from the left) the point spread function dropped significantly as was also expected from the recorded pictures (see figures \ref{fig:before} and \ref{fig:after}). In the beginning of the year 2016 the PSF degraded very suddenly and it was most probably caused by the high wind speeds and snow. The dimmer the datapoint, the less important the night was due to less muons detected.

            As one might notice when comparing figures \ref{fig:hough_psf_response} and \ref{fig:after}, the measured point spread function is much larger than the reconstructed one after the mirror alignment on 14th of May 2014. The difference probably comes from the fact, that if even one mirror is misaligned a lot then the image cleaning will not find the photons reflected by this particular facet because it will not be in a dense enough cluster. In contrast, the equipment measuring point spread function is able to spot also the photons reflected by the misaligned facet. This is however not a big shortcoming of the reconstruction of the image cleaning because it takes into account only the differences, not the overall value, thus it is possible to scale it accordingly if PSF at one point is known.

\newpage
\chapter{\label{sec:Concl}Conclusion and Outlook}

    \section{Conclusion}

        Finding muon rings and using them to infer the true optical point spread function turned out to be a good method that can be used efficiently also on FACT.

        When using photon stream, using density clustering (in our case DBSCAN) turned out to be very efficient and precise method for identifying Cherenkov photons. We managed to even increase the precision by making cuts to get rid of the branch-like structures which emerged due to inhomogeneities in the night-sky-background.

        After finding the Cherenkov photons, an accurate way to reconstruct the muon ring was needed. The most accurate way was found out to be DB-RM-HT, which made use of scikit-learn's CircleModel by taking it's results as a rough estimate for the ring parameters. These ring parameters were used as an initial guess for Hough transform. In contrast with AstroPy's Circle2D model, our Hough transform used triangular weighing function instead of the Heaviside function.

        Third obstacle to overcome was finding the best parameter for the fuzziness of the muon ring, which then could be used as a tool to infer the true optical point spread function. The fuzziness parameter 'response' turned out to be the best due to it's smaller uncertainty.

        As a final step the fuzziness parameter was to be calculated using real observations. After getting the fuzziness over time it was possible to infer the point spread function fit that we got by fitting the datapoints that we got from simulating different night-sky-backgrounds.

        It was clearly visible from the final PSF-vs-time plot how the reconstructed PSF correlated very well with the mirror alignments. Due to stormy weather in the beginning of 2016 some of the mirror facets were misaligned thus worsening the point spread function, which is clearly visible from the presented figures. Using the results it is clear that mirrors should be again aligned. 

        All these found methods turned out to be very good when one wants to infer the PSF from the past, thus allowing one to reconstruct gamma-ray flux with better instrument response functions.

    \section{Outlook}

        Although found methods were good in achieving the goals, there is still room to improve the given methods, that were not implemented in this thesis.

        Finding Cherenkov photons from the night-sky-background photons might benefit when one would adjust DBSCAN's $\epsilon$-neighborhood depending on the night-sky-background rate. Also the cut made after DBSCAN can be adjusted depending on the night-sky-background rate.

        After finding the Cherenkov photons and reconstructing the rings one might want to find better cut values for classifying events to be muon-like or not, thus increasing the sensitivity and precision of the detection method. For the evaluation of the detection method a bigger manually classified set of events would be beneficial. This can be achieved by using Zooniverse as did VERITAS (see \cite{zooniverse}).

\newpage
\chapter*{Acknowledgements}
    I wish to thank my supervisor prof. Dr. Adrian Biland, who read through my drafts multiple times and gave very good advice on how to continue my thesis. Furthermore Sebastian A. Müller and Dominik Neise were always there to give advice and help out in a difficult situation. Last but not least Axel Arbet-Engels was very helpful giving comments and opinions when I needed them. 

\newpage

\bibliography{ThesisPaper}{}

\newpage

\begin{appendices}
\chapter{Other muon ring feature extraction methods}
    \section{DB-RM-SD} \label{SD}
    In hopes that ring center position is reconstructed accurately enough, method DB-RM-SD was tested.

    Once again in order to identify Cherenkov photons, density clustering is used. Now instead of directly using the results from RingModel, we use the found ring-center positions and estimate ring radius once again by calculating the distance of every single Cherenkov photon from the previously found center and taking the median of the distances to be the new radius.

\begin{figure}[H]
            \begin{center}
                \includegraphics[width=0.7\textwidth]{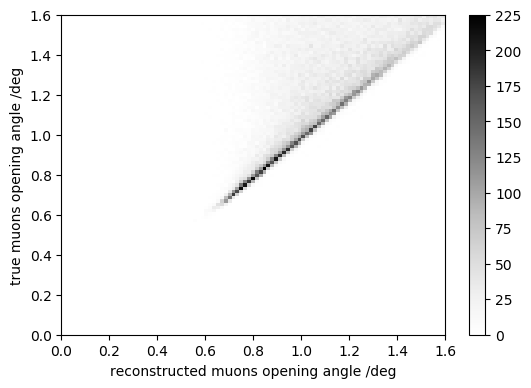}
                \caption{Confusion matrix for opening angle with DB-RM-SD method}
                \label{fig:med_r_oa}
            \end{center}
        \end{figure}

    As one can see figure \ref{fig:med_r_oa}, then the reconstructed radius still had very similar characteristic - it once again underestimated the ring radius very often.

\begin{figure}[H]
            \begin{center}
                \includegraphics[width=0.7\textwidth]{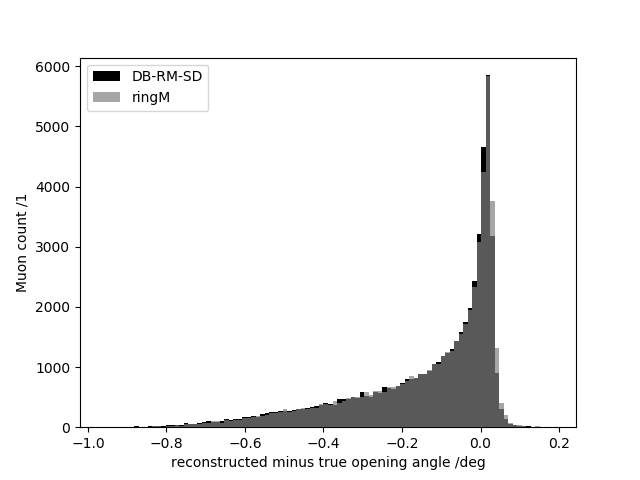}
                \caption{Reconstructed opening angle minus true opening angle}
                \label{fig:medR_oa}
            \end{center}
        \end{figure}

    When subtracting true opening angle from reconstructed one one sees how bad the radius is reconstructed (see figure \ref{fig:medR_oa}). The standard deviation of the difference of reconstructed opening angle and true opening angle for this method is even worse than before 0.187 degrees. So, this method doesn't imrove the results from DB-RM.

    \section{DB-RM-KC} \label{KC}

\begin{figure}[H]
            \begin{center}
                \includegraphics[width=0.7\textwidth]{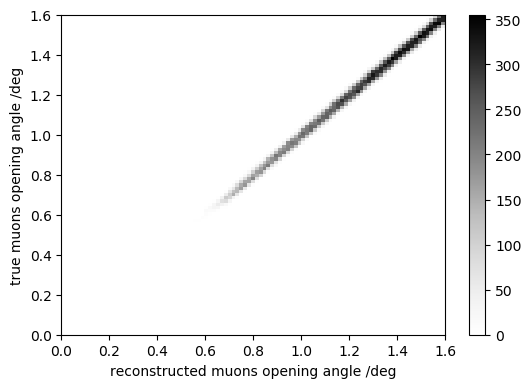}
                \caption{Confusion matrix for opening angle with known center}
                \label{fig:k_c_r}
            \end{center}
        \end{figure}

    Inaccuracy in finding correct ring center motivated to test the case when ring center was as accurate as possible - it was taken from the simulation truth. Although one cannot know the true ring center in reality, this check showed that guessing accurately the ring center has a huge impact in reconstructing the ring radius. As one can see in figure \ref{fig:k_c_r} then in that case the muon ring radius was also reconstructed very accurately as one would have expected. Of course this was only a test how much the choice of the center position influences the reconstrucion of the ring radius.

\begin{figure}[H]
            \begin{center}
                \includegraphics[width=0.7\textwidth]{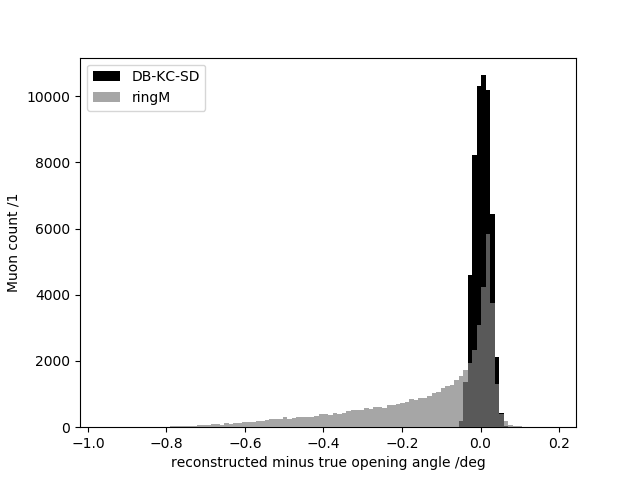}
                \caption{Reconstructed opening angle of the Cherenkov cone}
                \label{fig:knownC_oa}
            \end{center}
        \end{figure}

    As one can see from figure \ref{fig:knownC_oa} the ring radius is reconstructed very accurately. The reason for this not being exactly one is probably the pixellation of the camera. The standard deviation of the difference of extracted opening angle and true opening angle is 0.0202 degrees. This method showed that the ring center and radius from only using circle model from skimage is not good enough even though ring center seemed to be reconstructed well.

\chapter{Found distributions opening angle}

\begin{figure}[H]
        \begin{center}
            \includegraphics[width=\linewidth]{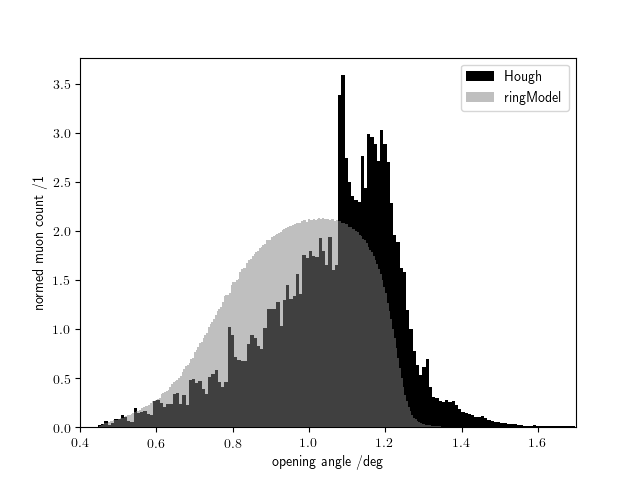}
            \caption{Comparison of two models opening angle distribution. Both distributions are normed}
            \label{fig:real_comaparison_oa}
        \end{center}
    \end{figure}

    As seen before from figure \ref{fig:circleHough_oa} and \ref{fig:hough_oa} we see that Hough Transform reconstructs ring parameters much more accurately than using the ringM method. Comparing the distributions on real data (figure \ref{fig:real_comaparison_oa}), we can once again see, that the distributions are different - ringM method reconstructs rings to be with smaller radii while with the added Hough transform the reconstructed radius is shifted towards the bigger radii. 

    Furthermore, from figure \ref{fig:real_comaparison_oa} we can see that the reconstructed ring radius for Hough transform peaks at opening angle $\sim 1.22$ degrees, which was also expected (see \ref{eq:openingAngle}). For ringM it peaks at $\sim 1$ degree.

    The reason for Hough transform (DB-RM-HT) not being as smoothly distributed as for DB-RM is most probably because of binning effects when performing the Hough transform.

\end{appendices}
\end{document}